# Neutron diffraction study of spin and charge ordering in SrFeO$_{3-\delta}$

M. Reehuis, [1] C. Ulrich, [2,3,4] A. Maljuk, [2,5] Ch. Niedermayer, [6] B. Ouladdiaf, [7] A. Hoser, [1] T. Hofmann, [1] and B. Keimer [2]

[1] *Helmholtz-Zentrum Berlin für Materialien und Energie, D-14109 Berlin, Germany*
[2] *Max Planck Institute for Solid State Research, Heisenbergstr. 1, D-70569 Stuttgart, Germany*
[3] *University of New South Wales, School of Physics, Sydney, NSW 2052, Australia*
[4] *Australian Nuclear Science and Technology Organisation (ANSTO), Locked Bag 2001, Kirrawee DC, NSW 2232, Australia*
[5] *Leibniz Institut für Festkörper- und Werkstoffforschung (IFW) Dresden, Helmholtzstr. 20, D-01171 Dresden, Germany*
[6] *Labor für Neutronenstreuung, Paul-Scherrer-Institut, CH-5232 Villigen, Switzerland*
[7] *Institut Laue-Langevin, BP 156, F-38042 Grenoble Cedex 9, France*

We report a comprehensive neutron diffraction study of the crystal structure and magnetic order in a series of single-crystal and powder samples of SrFeO$_{3-\delta}$ in the vacancy range $0 \leq \delta \leq 0.23$. The data provide detailed insights into the interplay between the oxygen vacancy order and the magnetic structure of this system. In particular, a crystallographic analysis of data on Sr$_8$Fe$_8$O$_{23}$ revealed a structural transition between the high-temperature tetragonal and a low-temperature monoclinic phase with a critical temperature $T = 75$ K, which originates from charge ordering on the Fe sublattice and is associated with a metal-insulator transition. Our experiments also revealed a total of seven different magnetic structures of SrFeO$_{3-\delta}$ in this range of $\delta$, only two of which (namely an incommensurate helix state in SrFeO$_3$ and a commensurate, collinear antiferromagnetic state in Sr$_4$Fe$_4$O$_{11}$) had been identified previously. We present a detailed refinement of some of the magnetic ordering patterns and discuss the relationship between the magneto-transport properties of SrFeO$_{3-\delta}$ samples and their phase composition and magnetic microstructure.

PACS number(s): 61.05.fm, 61.66.Fn, 75.25.-j, 75.50.E

# I. INTRODUCTION

The physical properties of manganese oxides have been the focus of much research activity over the past two decades, with particular emphasis on the "colossal magnetoresistance" effect. The compound $SrFeO_3$ crystallizes in a perovskite structure closely similar to the one of $LaMnO_3$, the progenitor of one of the most extensively studied manganate families. Moreover, the $Fe^{4+}$ ions in $SrFeO_3$ exhibit the same high-spin $3d^4$ electron configuration as the $Mn^{3+}$ ions in $LaMnO_3$. Despite these similarities, the electronic phase behavior of these two materials is completely different. While $LaMnO_3$ is a Mott insulator and shows orbital order and commensurate, collinear antiferromagnetism, $SrFeO_3$ is metallic and orbitally degenerate at all temperatures and exhibits a transition to incommensurate, helical magnetic order at $T_N = 134$ K.[1] The profound difference between the electronic properties of both compounds is believed to originate from the higher degree of covalency of the metal-oxide bond in $SrFeO_3$, but a detailed microscopic model picture has not yet been developed. This is in part due to the complex materials chemistry of $SrFeO_{3-\delta}$, which can be prepared in stoichiometric form ($\delta \sim 0$) only under high oxygen pressure. Samples with $\delta > 0$ exhibit a mixture of structural phases with different ordering patterns of oxygen ions, which go along with different arrangements of iron ions in square-pyramidal $FeO_5$ and octahedral $FeO_6$ coordination.[2-6] In addition to the stoichiometric end member $SrFeO_3$, which crystallizes in the cubic perovskite structure (space group $Pm\bar{3}m$), these include tetragonal $Sr_8Fe_8O_{23}$ (space group $I4/mmm$), orthorhombic $Sr_4Fe_4O_{11}$ (space group $Cmmm$), and orthorhombic, incommensurately modulated $Sr_2Fe_2O_5$ (superspace group $I2/m(0\beta\gamma)0s$).[6-9] The crystal structures of the former three compounds, which are subjects of the current article, are shown in Fig. 1.

Prior work has led to the identification of three separate magnetoresistance effects in $SrFeO_{3-\delta}$ with $\delta \leq 0.25$ (Refs. 10-14), which can be traced to an additional magnetic transition below the Néel temperature in $SrFeO_3$ (Refs. 11 and 12), to a charge-ordering transition in $Sr_8Fe_8O_{23}$, (Refs. 11 and 12), and to magnetic disorder in $Sr_4Fe_4O_{11}$ (Refs. 10-14), respectively. The coexistence between different oxygen vacancy-ordered phases in samples with $\delta \neq 0$ has so far confounded the quantitative description of these effects, because the superposition of different diffraction patterns in mixed-phase samples greatly complicates the crystallographic identification and analysis of the underlying spin and charge ordering patterns. At the same time, real-space observations of oxygen vacancy order by electron microscopy are difficult, because these structures are very susceptible to electron beam damage.[15]

Here we present a neutron diffraction study designed to address this problem. By virtue of a detailed crystallographic analysis of single-crystal and powder samples of $SrFeO_{3-\delta}$ with different concentrations of oxygen vacancies, we have arrived at a detailed experimental description of the spin and charge ordering patterns in this family of compounds. Specifically, we report a comprehensive determination of the crystal structure of $Sr_8Fe_8O_{23}$, which yields insight into the charge ordering pattern at low temperature. In addition, we pinpoint the origin of the transport anomaly in nominally stoichiometric crystals of cubic $SrFeO_3$ as a hitherto unidentified commensurate magnetic phase, whose onset strongly affects the domain structure of the helical majority phase. In the range $0 \leq \delta \leq 0.25$, we have identified a total of seven magnetic phases, only two of which – namely the incommensurate helical phase in $SrFeO_3$, and a commensurate collinear phase in $Sr_4Fe_4O_{11}$ – had been described before.[1,8] Among the newly identified phases are commensurate and incommensurate magnetic structures in $Sr_8Fe_8O_{23}$ as well as several weak magnetic structures that cannot be uniquely associated with any of the previously identified vacancy-ordered phases. This implies that the magnetic phase diagram of $SrFeO_{3-\delta}$ is much richer than previously believed. The plethora of magnetic phases generated by oxygen vacancies in $SrFeO_{3-\delta}$ is interesting in the context of recent work on magnetic field induced transitions and high-field magnetoresistance in $Sr(Fe,Co)O_3$ (Ref. 16), as well as related phenomena in other incommensurate magnets such as $BiFeO_3$ (Ref. 17), MnSi (Ref. 18), and MnGe (Ref. 19).

## II. EXPERIMENTAL DETAILS

High-quality oxygen-deficient single crystals of $SrFeO_{3-\delta}$ with volumes of up to 0.25 $cm^3$ were grown using the crucible-free floating zone technique as described elsewhere.[20] The oxygen-deficiency depends on the growth conditions. Under a moderate oxygen-pressure between 0.2 and 3.0 bar and a cooling rate 20-30 °C/h, an oxygen deficiency of $\delta = 0.23 \pm 0.02$ was obtained. The oxygen content in single crystals was further increased ($0 < \delta < 0.09$) under a post-growth high oxygen pressure treatment at pressures up to 700 bar.[21] In order to obtain the full stoichiometry of the end member $SrFeO_3$, the single crystals were annealed at 5 kbar oxygen pressure and a temperature of 400 °C.[21] In addition, we have investigated powder samples of $SrFeO_{3-\delta}$ with different oxygen contents. For the present study we annealed two powder samples at 400° C in oxygen flow and in air, reaching oxygen deficiencies of $\delta = 0.13 \pm 0.02$ and $\delta = 0.19 \pm 0.02$, respectively. The oxygen content in all samples was determined by thermogravimetry.

Single crystals of $SrFeO_{3-\delta}$ ($0 \leq \delta \leq 0.23$) were investigated on the 4-circle diffractometers E5 at the BER II reactor of the Helmholtz-Zentrum Berlin, and D10 at the Institut Laue-Langevin in Grenoble. Both instruments use pyrolytic-graphite (PG) monochromators selecting the neutron wavelength $\lambda = 2.36$ Å. Additional neutron diffraction experiments were performed on the triple-axis instrument RITA II at the Swiss continuous neutron spallation source (SINQ) at the Paul-Scherrer Institute in Villigen, Switzerland, where the same neutron wavelength of $\lambda = 2.36$ Å was used. This triple-axis instrument was used to measure selected Bragg reflections with higher accuracy and a significantly improved background. Complementary powder diffraction patterns of $SrFeO_{3-\delta}$ with $\delta = 0.13$ and $\delta = 0.19$ were recorded on the powder diffractometer E6 at the BER II reactor of the Helmholtz-Zentrum Berlin, where the neutron wavelength $\lambda = 2.44$ Å was selected by a PG-monochromator. Additional powder patterns of $SrFeO_{3-\delta}$ with $\delta = 0.13$ were recorded on the high-intensity and high-resolution neutron powder diffractometers DMC and HRPT at the Paul-Scherrer-Institut. The instruments DMC and HRPT used a PG- and a Ge-monochromator selecting the neutron wavelengths $\lambda = 2.57$ Å and $\lambda = 1.158$ Å, respectively. The refinements of the crystal and magnetic structure were carried out with the program *FullProf* (Ref. 22) with the nuclear scattering lengths $b(O) = 5.805$ fm, $b(Fe) = 9.54$ fm, and $b(Sr) = 7.02$ fm.[23] The magnetic form factors of the $Fe^{3+}$- and $Fe^{4+}$-ions were taken from Ref. 24.

### III. RESULTS AND DISCUSSION

**A. Coexistence of vacancy-ordered and magnetic phases in $SrFeO_{3-\delta}$**

We begin our presentation by describing the determination of the structural and magnetic phase composition of our $SrFeO_{3-\delta}$ samples. To this end, we measured the temperature dependence of various nuclear and magnetic Bragg reflections from single crystals with oxygen deficiencies of $\delta = 0.0$, $0.03$, $0.13$, and $0.23$. Figure 1 shows that the iron sublattices in the crystal structures of $SrFeO_3$, $Sr_8Fe_8O_{23}$, and $Sr_4Fe_4O_{11}$ are very similar. This implies that the strong nuclear reflections appear at almost the same $2\theta$-positions, although the coordination of some fraction of the iron atoms changes from octahedral ($FeO_6$) to square-pyramidal ($FeO_5$), and the associated lattice distortions lower the space-group symmetry. Therefore, we give the indices of the measured reflections ($h, k, \ell$) in the cubic setting, in addition to the tetragonal and/or orthorhombic settings that characterize the full lattice symmetry of the oxygen-deficient compounds.

Figure 2 shows typical neutron diffraction data on the $SrFeO_{3.00}$ and $SrFeO_{2.87}$ single crystals. The measurements were performed at various temperatures along the [1,1,1] direction around the structural Bragg reflection $(0, 0, 1)_{cub}$. We first discuss the data on the crystal with the ideal composition $SrFeO_{3.00}$, which served as calibration standard for the determination of the phase fractions of the other samples. The $SrFeO_{3.00}$ crystal exhibits only one set of magnetic satellite peaks at the incommensurate wave vector ($\Delta, \Delta, \Delta$) with $\Delta$ = 0.129 at low temperatures. This signal can be attributed to the previously identified helical order of the iron moments (termed "phase I" in the present work) in the cubic structure of $SrFeO_3$.[1] The temperature dependence of the corresponding magnetic Bragg peak intensity is displayed in Fig. 3. The Bragg reflections of phase I vanish upon heating above the Néel temperature $T_N$ = 133(1) K, which is in excellent agreement with $T_N$ = 134 K given earlier.[1]

The diffraction data of the $SrFeO_{2.87}$ crystal, also presented in Fig. 2, are characteristic of mixed-phase samples. In addition to the phase-I reflections, this sample exhibits a second set of magnetic reflections with $\Delta$ = 0.20, which disappear for temperatures above $T_S$ = 75(2) K. The onset temperature of the magnetic structure indicated by these reflections (henceforth termed "phase II") coincides with the onset of magnetic and charge order previously observed by Mössbauer spectroscopy, transport, optical, and Raman scattering experiments in the tetragonal ferrate $Sr_8Fe_8O_{23}$ (Refs. 11 and 12). The presence of a structural phase transition due to charge ordering is confirmed by an intensity change of the nuclear reflection $(2, 0, 0)_{cub}$ [$(4, 4, 0)_{tetr} / (0, 0, 4)_{tetr}$] at $T_S$ (Fig. 4). The main cause of the intensity anomalies of the main nuclear Bragg reflections at the charge ordering transition is a modification of the mosaic character of the single crystals due to lattice strains and a consequent change in the extinction of the neutron beam. A detailed crystallographic analysis (see Section III.B) indicates a small additional contribution from atomic displacements associated with charge ordering.

Figure 4 also displays corresponding data for the crystal of composition $SrFeO_{2.97}$, where the anomaly of the $(2, 0, 0)_{cub}$ reflection is reduced compared to the one in the $SrFeO_{2.87}$ crystal, and the intensity of the magnetic satellite at $(0.129, 0.871, 0.871)_{cub}$ is correspondingly enhanced. The transition temperatures ($T_S$ = 75 and $T_N$ = 133 K, respectively), on the other hand, do not depend on oxygen content. This indicates that these crystals contain different fractions of the cubic ($SrFeO_3$) and vacancy-ordered tetragonal ($Sr_8Fe_8O_{23}$) phases. Twinning of the crystal structures of the lower-symmetric ferrates and the presence of various minority phases (see below) greatly complicate the full refinement of the lattice and magnetic structures in the single crystals with $\delta > 0$. In order estimate the phase composition of these samples, we have therefore used the helical magnetic structure of $SrFeO_3$, whose ordered moment [$\mu_{exp}$ = 2.960(12) $\mu_B$ per $Fe^{4+}$-ion] has been precisely determined at 2 K (see Section III.C.1), in order to determine the overall scale factor of $SrFeO_3$ from the intensity of the incommensurate magnetic reflection (0.129, 0.871, 0.871)$_{cub}$. In this way, we found that the two samples $SrFeO_{2.97}$ and $SrFeO_{2.87}$ contain 74(2) and 26(2) % of the cubic phase $SrFeO_3$, respectively. The other main component in these samples is the tetragonal phase $Sr_8Fe_8O_{23}$.

The oxygen content of the $SrFeO_{2.77}$ crystal is close to that of the orthorhombic phase $Sr_4Fe_4O_{11}$ ($SrFeO_{2.75}$). In fact, magnetic intensity could be found on the position of the magnetic reflection (1, 1, 1)$_{orth}$, suggesting the presence of the previously identified commensurate magnetic structure of $Sr_4Fe_4O_{11}$ (here termed "phase VII"). Fig. 5 shows that the magnetic intensity vanishes at the Néel temperature $T_N$ = 232(2) K, which is in excellent agreement with the transition temperature given earlier.[8]

A comprehensive survey of the reciprocal lattice of our single-crystal samples revealed several sets of weaker magnetic reflections, which could not be uniquely associated with one of the structures displayed in Fig. 1 and may hence indicate small fractions of hitherto unidentified oxygen-vacancy ordered structures. In the crystals with composition $SrFeO_{3.00}$, $SrFeO_{2.87}$, and $SrFeO_{2.77}$, we observed magnetic reflections that could be indexed as (¼, 0, 0)$_{cub}$ and (¾, 0, 0)$_{cub}$, or (0, 0, ½)$_{tetr}$ and (0, 0, 1½)$_{tetr}$ in the tetragonal setting, below $T_N$ = 65(4) K (Figs. 3 and 5). Their intensity in the $SrFeO_{3.00}$ crystal is much weaker than in the other crystals, and may be due to residual oxygen defects below the detection level of 0.02 for $\delta$ in $SrFeO_{3-\delta}$. The onset temperature of these reflections coincides with anomalies in Mössbauer and transport measurements previously reported in nominally stoichiometric $SrFeO_{3-\delta}$ crystals.[12] They can be generated with the propagation vector $\boldsymbol{k}$ = (0, 0, ½), indicating a magnetic unit cell where the tetragonal $c$-axis is doubled. We refer to this structure as "phase IV" and discuss its properties in detail in Section III.C.4.

In the $SrFeO_{2.77}$ crystal, very weak magnetic intensity could be observed at the position (0.30, 0.30, 0.75)$_{cub}$. The peak intensity disappears upon heating above $T$ = 110(4) K, which coincides with anomalies in susceptibility and resistivity measurements on samples with different oxygen contents (Fig. 5).[12] Furthermore, incommensurate magnetic Bragg reflections were observed at $(0.79, 0.79, 0)_{cub}$ and $(1.21, 1.21, 0)_{cub}$ which disappear at a transition temperature of $T = 60(5)$ K. The incommensurate magnetic phases V and VI indicated by these two sets of reflections may be attributed to small volumes of hitherto unidentified oxygen-deficient ferrates with crystal structures very similar to those of $Sr_8Fe_8O_{23}$ and $Sr_4Fe_4O_{11}$, which are intergrown in the majority-phase matrix. Due to the weakness of the observed reflections, the magnetic structures of phases V and VI could not be determined from the available data. However, a description of these structures in the tetragonal setting implies that incommensurate ordering occurs in the basal *ab*-plane.

The widths of the magnetic reflections are generally significantly larger than the instrumental resolution and exhibit interesting temperature dependences (Fig. 3). In the $SrFeO_{3.00}$ crystal, the widths of the incommensurate phase-I reflections at high temperatures indicates a magnetic domain size of ~100 Å (Fig. 3, left panel). The small domain size probably reflects phason disorder, along with a distribution of different polarization domains. Below the onset temperature of phase IV, however, the domain size inferred from the width of the phase-I reflections increases by a factor of two, possibly because the magnetic order in the phase-IV domains locks in the phase relationship between some of the phase-I domains. The strong influence of the onset of phase IV on the microstructure of phase I, despite the small volume fraction of phase IV, implies that both phases are not macroscopically separated, but rather densely intertwined. A possible scenario is that the small number of residual oxygen vacancies that generate phase IV form extended planar structures, as they do in $Sr_8Fe_8O_{23}$ and $Sr_4Fe_4O_{11}$, but with a different periodicity.

A related effect is observed in the $SrFeO_{2.87}$ crystal, where the size of the phase-I domains decreases below the onset temperature of phase II (Fig. 3, right panel). At the same temperature, the growth of the phase-I intensity upon cooling is arrested, possibly because some of the volume occupied by phase I for $T > T_S = 75$ K is converted to phase II. These data indicate that the interplay of the oxygen vacancy order and the magnetic structure generates a complex, nanoscopic magnetic landscape.

Complementary neutron diffraction experiments were performed on $SrFeO_{3-\delta}$ powder samples with oxygen deficiencies $\delta = 0.13$ and $\delta = 0.19$, in the temperature range between 10 and 240 K. The experiments were carried out on the instruments E6 and DMC using the larger neutron wavelengths $\lambda = 2.44$ Å and $\lambda = 2.57$ Å, respectively. The powder patterns of the sample with $\delta = 0.13$, presented in Fig. 6, clearly show the appearance of individual sets of magnetic reflections with different ordering temperatures. The strongest magnetic reflection, observed at $2\theta = 8.3°$, could be identified as the magnetic satellite $(0.128, 0.128, 0.128)_{cub}$ of the helical phase I, which appears below the Néel temperature $T_N = 133(1)$ K. The wave vector components were found to be closely similar to those obtained from our single-crystal diffraction study. A second set of four incommensurate magnetic reflections appears below $T_S = 75(2)$ K, the charge ordering temperature of $Sr_8Fe_8O_{23}$ already discussed above. However, the magnetic structure indicated by these reflections is different from the one identified in the same temperature range in the crystal with the same composition, $SrFeO_{2.87}$. We therefore refer to this structure as "phase III". (Note that reflections characteristic of phase III were also found in the single-crystalline sample, albeit with much weaker intensity.) The strongest phase-III satellite, observed at $2\theta = 11.2°$, could be indexed as $(0.169, 0.169, 0.169)_{cub}$ using the cubic setting. In contrast, the other three satellites of this set observed at $2\theta$-angles of 13.6°, 18.9°, and 29.6° (Fig. 6) could not be generated by the rule $(hk\ell)_M = (hk\ell)_N \pm \boldsymbol{k}$ using the cubic unit cell. Taking into account the much larger unit cell of the tetragonal phase $Sr_8Fe_8O_{23}$ (cell dimensions $2\sqrt{2}a \times 2\sqrt{2}a \times 2c$ referenced to the cubic cell), all of these satellites could be indexed. Here the cubic cell has been transformed to the tetragonal unit cell via the matrix (220; 2–20; 002). Consequently the propagation vector $\boldsymbol{k} = (0.169, 0.169, 0.169)_{cub}$ changes to $\boldsymbol{k} = (0.676, 0, 0.338)_{tetr}$, where the $\boldsymbol{k}$-vector now lies in the tetragonal $ac$-plane. Finally, magnetic reflections indicative of phase IV were detected at $2\theta = 9.5°$ and $2\theta = 28.9°$ (Fig. 6) below 65(4) K. In the cubic setting these two reflections could be indexed as $(¼, 0, 0)_{cub}$ and $(¾, 0, 0)_{cub}$, [or $(0, 0, ½)_{tetr}$ and $(0, 0, 1½)_{tetr}$], as discussed above for the single crystal data.

In the powder sample with oxygen deficiency $\delta = 0.19$, the cubic ferrate $SrFeO_3$ was found to be absent, but magnetic reflections characteristic of the orthorhombic ferrate $Sr_4Fe_4O_{11}$ were observed. With the knowledge of the magnetic structures of the different ferrates, we were able to determine the composition of the two powder samples (see Section III.C.4)

### B. Crystal structure and charge ordering transition of $Sr_8Fe_8O_{23}$

The data presented in the previous section demonstrate that all of the oxygen-deficient single-crystalline samples contained at least three different vacancy-ordered structures. The lower-symmetric structures show twinning, and the tetragonal ferrate $Sr_8Fe_8O_{23}$ exhibits additional twinning below the low-temperature structural phase transition associated with charge ordering. Therefore it was not possible to refine the crystal structures of the oxygen-deficient ferrates from the single-crystal diffraction data. In order to investigate the crystal structure of $Sr_8Fe_8O_{23}$ ($SrFeO_{2.875}$) we thus collected high-resolution neutron powder patterns (Fig. 7) using a sample with composition $SrFeO_{2.87}$ which mostly contains $Sr_8Fe_8O_{23}$. The choice of the shorter neutron wavelength $\lambda = 1.158$ Å allowed us to collect a relatively large number of nuclear Bragg reflections up to $\sin\theta/\lambda = 0.85$ Å$^{-1}$. Rietveld refinements of the powder data collected at 199 K in the tetragonal space group *I*4/*mmm* (No. 139) resulted in a residual $R_F = 0.038$ (defined as $R_F = \sum||F_{obs}| - |F_{calc}||/\sum|F_{obs}|$) and a reduced $\chi^2 = 1.45$. The crystal structure of the minor phase $SrFeO_3$ was refined simultaneously. In agreement with Hodges *et al.* (Ref. 6) we could not obtain a better fit using the orthorhombic and monoclinic space groups *Fmmm* ($R_F = 0.055$, $\chi^2 = 2.01$) and *I*2/*m* (setting: *I* 1 1 2/*m*) ($R_F = 0.058$, $\chi^2 = 2.05$), respectively. The Sr-, Fe-, and O-atoms in *I*4/*mmm* are located at the following Wyckoff positions: Sr1 in 8*i*(*x*,0,0), Sr2 in 8*j*(*x*,0,½), Fe1 in 4*e*(0,0,*z*), Fe2 in 8*f*(¼,¼,¼), Fe3 in 4*d*(½,0,¼), O1 in 2*b*(0,0,½), O2 in 16*m*(*x*,*x*,*z*), O3 in 8*h*(*x*,*x*,½), O4 in 16*k*(*x*,*x*+½,¼), and O5 in 4*c*(½,0,0).

In Fig. 1 the crystal structure of $Sr_8Fe_8O_{23}$ is compared with the structures of the cubic and orthorhombic phases $SrFeO_3$ and $Sr_4Fe_4O_{11}$, respectively. $Sr_4Fe_4O_{11}$ (space group *Cmmm*) is characterized by corner-sharing square pyramids (FeO$_5$) along the orthorhombic *c*-axis. Due to the oxygen deficiency at one of the apical oxygen sites, the Fe-atom at 4*i*(0,*y*,0) is shifted out of the plane [or from the ideal value $y = ¼$ to 0.253(1)], which is formed by the four equatorial oxygen atoms. This can be attributed to the strong bond between the iron atom and the apical atom O1, which is fixed at the position 2*b*(0,½,0) in *Cmmm*.[6,8] Note that the *b*-axis of the orthorhombic setting corresponds to the *c*-axis in the tetragonal one (Fig. 1). The same behavior was observed for the FeO$_5$-units in $Sr_8Fe_8O_{23}$, where the *z*-parameter of the Fe1-atom [at 4*e*(0,0,*z*)] in the FeO$_5$-unit is shifted from the ideal value $z = ¼$ to 0.2546(6) at 90 K and 0.2543(6) at 199 K, respectively. This Fe1-atom is then also closer to the apical O1-atom, which is fixed at the position 2*b*(0,0,½) in *I*4/*mmm* (Table I). Our results disagree with a previous report on the room-temperature structure of $Sr_8Fe_8O_{23}$, which assigned the ideal position of 0.250(2) to the Fe1-atom.[6] This result is surprising, because our refinements showed that the *z*-parameter of Fe1 does not change significantly between 90 and 199 K (Table I). The equatorial oxygen atoms in both $Sr_4Fe_4O_{11}$ and $Sr_8Fe_8O_{23}$ show shifts in the direction opposite to one of the iron atoms: O3 [16*r*(*x*,*y*,*z*)] in $Sr_4Fe_4O_{11}$ from $y = ¼$ to 0.2239(1); O2 [16*m*(*x*,*x*,*z*)] in $Sr_8Fe_8O_{23}$ from $z = ¼$ to 0.2227(4) at 90 K and 0.2218(4) at 199 K. These distortions of the square-pyramidal $FeO_5$-units induce tilts of the corner-shared $FeO_6$-octahedra, as shown in Fig. 1.

For a detailed investigation of the structural modifications associated with the charge-ordering transition at $T_S = 75$ K, we used powder patterns collected at 2 and 90 K (Fig. 7). The crystal structure of $Sr_8Fe_8O_{23}$ at 90 K could be again refined in the space group $I4/mmm$, resulting in a residual $R_F = 0.034$ ($\chi^2 = 1.45$), which is similar to that one obtained from the data set collected at 199 K. The refinement of the structure at 2 K in the tetragonal space group resulted in a similar residual $R_F = 0.037$, but in a considerably larger $\chi^2$-value ($\chi^2 = 2.24$). In contrast to the powder pattern collected at 90 K, it could be seen that some of the refined peak positions in the $2\theta$-range above 130° differed from the observed positions. As shown in the inset of Fig. 7, the $2\theta$-positions of reflections ($h, k, \ell$) with large $\ell$ showed the strongest discrepancies. This led us to refine the crystal structure of $Sr_8Fe_8O_{23}$ in a lower-symmetric space group. Among all the space groups that could be deduced from group-subgroup relations, the best fit was obtained for the monoclinic space group $I2/m$ (No. 12, setting $I 1 2/m 1$, standard setting $C2/m$) resulting in a residual $R_F = 0.037$ ($\chi^2 = 1.92$). The observed and calculated powder patterns at 2 K and 90 K are shown in Fig. 7.

Due to the symmetry reduction, the Wyckoff positions of the atoms Sr1, Sr2, Fe2, O2, O3, O4 and O5 in $I4/mmm$ split into two different sites: Sr11 in $4i(x,0,z)$ and Sr12 in $4g(0,y,0)$, Sr21 in $4i(x,0,z)$ and Sr22 in $4h(0,y,½)$, Fe21 in $4e(¼,¼,¼)$ and Fe22 in $4f(¼,¼,¾)$, O21 and O22 as well as O41 and O42 in $8j(x,y,z)$, O51 in $2b(0,½,0)$ and O52 in $2c(½,0,0)$. For the atoms Fe1, Fe3 and O3 the multiplicities are retained: Fe1 and Fe3 in $4i(x,0,z)$, O3 in $8j(x,y,z)$. In the tetragonal structure ($I4/mmm$) a total number of 7 positional parameters were refined, while in the lower symmetric space group $I2/m$ a much larger number of 25 positional parameters were allowed to vary. This was problematic within $I2/m$, as the parameters of the monoclinic phase are highly correlated. As a consequence, the parameters of the monoclinic structure could not be determined as accurately as those of the tetragonal phase (Table I). However, the refinements showed that the $z$-parameter of Fe3 is practically the same as the ideal value ¼ that is fixed in $I4/mmm$. Further the $x$- and $y$-parameters of O3 were found to be identical. Therefore in the last refinement the value $z$(Fe3) was fixed at ¼, and $x$(O3) and $y$(O3) were constrained to be equal.

The positional parameters and the distances between the oxygen and iron atoms obtained at 2, 90, and 199 K are listed in Tables I and II, and the $ab$-plane projections of the high-temperature tetragonal and the low-temperature monoclinic structures of $Sr_8Fe_8O_{23}$ are shown in Fig. 8. The octahedrally coordinated Fe2-atoms are tilted in the tetragonal phase due the oxygen vacancies at the Fe1-sites. Due to the tetragonal symmetry, the apical O3-atoms [in $8h(x,x,½)$] are shifted along the directions [110] and [1−10] or inverse. In the monoclinic structure, the independent parameters $x$ and $y$ of the O3-atoms [in $8j(x,y,z)$] did not show any significant shift from the positions obtained for the tetragonal phase. In Fig. 8 it can be seen that the tilts of the $FeO_6$-octahedra are not significantly influenced by the change of symmetry. The main changes in the monoclinic structure occur in the $ab$-plane. The O2-atoms in the tetragonal structure are located at the Wyckoff position O2 in $16m(x,x,z)$, while in the monoclinic setting this position is split into two positions O21 and O22, both in $8j(x,y,z)$. In Table I it can be seen that the refined parameters $x$(O21) and $y$(O22) as well as $y$(O21) and $x$(O22) are similar. Fig. 8 shows that the O21- and O22-atoms are shifted parallel to the directions [110] and [1−10] (or inverse) which are almost perpendicular to the Fe-O-bonds. On the other hand, the $x$- and $y$-parameters of the other equatorial O41- and O42-atoms were found to be similar.

Additional data sets were collected in the monoclinic phase at 30 and 60 K, as well as in the tetragonal phase at 119, 146, 168, and 227 K. The refinements showed that the structural parameters of the monoclinic phase up to 60 K are similar to those obtained from the powder pattern collected at 2 K. This shows that the monoclinic structure is already well established 15 K below $T_S$ = 75 K. From all these data sets we were able to determine the lattice parameters with good accuracy. In Fig. 9 it can be seen that the lattice parameters $a$ (= $b$) and $c$ show a continuous decrease in the tetragonal phase down to 90 K. Below the structural phase transition, only the $c$-parameter shows a significant change.

We now discuss the implications of these observations for our understanding of the charge ordering transition in $Sr_8Fe_8O_{23}$. At room temperature, this compound contains three different iron sites, namely Fe1 with square-pyramidal coordination and two crystallographically distinct atoms Fe2 and Fe3 with octahedral coordination, with a ratio of 1 : 2 : 1. Mössbauer spectra at room temperature showed two different components, $Fe^{3.5+}$ and $Fe^{4+}$, with a ratio of 1 : 1 (Ref. 11 and 12), and calculations of the Fe-O bond-strength sums around the Fe-sites indicated that the Fe2-ions possess a lower valence state than the Fe1- and Fe3-ions.[6] Based on this information, the $Fe^{4+}$ state could be assigned to both the Fe1- and the Fe3-atoms, while the Fe2-atoms are in the valence state $Fe^{3.5+}$.

We first focus on the behavior of the $Fe^{4+}$ ions. The almost regular $Fe3O_6$-octahedron in $Sr_8Fe_8O_{23}$ is reminiscent of the undistorted octahedra in cubic $SrFeO_3$, where the Fe-O bond distance is $d$(Fe-O) = 1.9255(5) Å (at room temperature). In orthorhombic $CaFeO_3$ the three different $d$(Fe-O) values vary between 1.9184(6) and 1.927(2) Å (averaged $d_{av}$(Fe-O) =

1.922(2) Å).[6,25] In tetragonal $Sr_8Fe_8O_{23}$ the averaged bond distance $d_{av}$(Fe3-O) = 1.916(2) Å (at room temperature) was found to be slightly smaller.[6] We found very similar values for the tetragonal phase $Sr_8Fe_8O_{23}$ at lower temperatures: $d_{av}$(Fe3-O) = 1.914(4) Å at 199 K and 1.913(4) Å at 90 K. In the monoclinic phase, neither the average Fe3-O distance [$d_{av}$(Fe3-O) = 1.917(7) Å at 2 K] nor the shape of the $Fe3O_6$-units differ significantly from the tetragonal phase. In agreement with prior Mössbauer spectroscopy results (Ref. 11 and 12), we therefore conclude that the Fe3-atoms do not participate in the charge ordering transition. The same conclusions hold for the Fe1-atoms, which are also in the $Fe^{4+}$ valence state. The average Fe1-O bond length in $Sr_8Fe_8O_{23}$, $d_{av}$(Fe1-O) = 1.864(8) Å at 2 K, is the same as that in the square-pyramidal $FeO_5$-units in $Sr_4Fe_4O_{11}$.[6,8] Again, only minor changes were observed at the charge ordering transition.

In contrast to the Fe1- and Fe3-atoms of $Sr_8Fe_8O_{23}$, the Fe2-O bonds are modified substantially at the charge-ordering transition. In the tetragonal charge-disordered phase, two of the six Fe2-O bonds in the $Fe2O_6$-octahedron are significantly elongated (Table II and Fig. 8), probably at least in part because of steric constraints imposed by adjacent oxygen vacancies. In the monoclinic charge-ordered phase, the Fe2-site splits into two different sites Fe21 and Fe22 with different averaged bond lengths $d_{av}$(Fe21-O) = 1.977(11) Å and $d_{av}$(Fe22-O) = 1.949(11) Å, which are larger and smaller than the value $d_{av}$(Fe22-O) = 1.962(3) Å obtained at 90 K, respectively. This is qualitatively consistent with a charge-ordering transition in the two-dimensional Fe2-atom network, in agreement with the conclusions from Mössbauer spectroscopy.[11,12] However, the difference in the averaged bond lengths of Fe21- and Fe22-atoms is only about half of the value $\Delta d$ = 0.055 Å, the difference of the bond distances of octahedrally coordinated Fe-atoms in the valence states $Fe^{3+}$ and $Fe^{4+}$ given by Shannon.[26] Note also that the difference in averaged bond lengths of the Fe-atoms with formal valence $Fe^{3+}$ and $Fe^{5+}$ in the monoclinic, charge-disproportionated phase of $CaFeO_3$ is even larger: $\Delta d$ = 0.102 Å.[25] This comparison indicates incomplete charge segregation in the monoclinic phase of $Sr_8Fe_8O_{23}$. Similar conclusions have been reported for other charge-ordered transition metal oxides, including in particular the charge-disproportionated state of $La_{1/3}Sr_{2/3}FeO_3$.[27] We also note that the Fe2-O bond-length modification in the charge-ordered state is anisotropic. In particular, only the two Fe21-O21 bonds increase significantly below the transition, while the other four bond lengths [$d$(Fe21-O41) and $d$(Fe21-O3)] in the $(Fe21)O_6$-octahedron are almost identical (Table II). In contrast, the elongated Fe22-O22-bonds do not show any significant change between the tetragonal and monoclinic structures. This indicates that charge ordering is associated with some degree of orbital polarization, which differs between the nominal $Fe^{3+}$ and $Fe^{4+}$ valence states, as expected. Detailed electronic-structure calculations will be required to explain the specific bond-distortion pattern we have observed.

### C. Magnetic order in $SrFeO_{3-\delta}$

#### *1. Phase I*

We now discuss some of the magnetic states identified in Section III.A in more detail. We begin with the stoichiometric end member $SrFeO_3$, where helical magnetic order with a propagation vector $\boldsymbol{k} = (\Delta, \Delta, \Delta)_{cub}$ and incommensurability $\Delta = 0.112$ was found earlier by Takeda *et al.*[1] In this study the magnetic moment of the iron atoms $\mu_{exp} = 2.7(4)$ $\mu_B$ was determined at the relatively high temperature of 77 K. In order to accurately determine the saturation value of the magnetic moment, we have reinvestigated the magnetic structure $SrFeO_3$ at $T = 2$ K by single-crystal neutron diffraction using the neutron diffractometer D10 at the ILL in Grenoble and the triple-axis spectrometer Rita-II at the PSI in Villigen/Switzerland. For the experiment we used a single crystal with the composition $SrFeO_{3.00}$ with dimensions $3 \times 6 \times 8$ mm$^3$. At the diffractometer D10 we first collected a set of 122 (9 unique) nuclear Bragg reflections in order to determine the overall scale factor from the crystal structure refinement, which is needed for the determination of the precise value of the ordered iron moment. In the cubic perovskite structure of $SrFeO_3$ with the space group $Pm\bar{3}m$ the Sr-, Fe-, and O-atoms are located at the following Wyckoff positions: Sr in $1b(½,½,½)$, Fe in $1a(0,0,0)$, O in $3d(0,0,½)$, where no positional parameters are refineable. Due to the fact that we only collected reflections up to $\sin\theta/\lambda = 0.39$ Å$^{-1}$ the temperature factors could not be determined with good accuracy. Therefore we used for the refinements a fixed value $B = 0.3$ Å$^2$ for all the atoms Sr, Fe, and O. The refinements of the overall scale factor and the extinction parameters resulted in a satisfactory residual of $R_F = 0.020$.

For the refinement of the magnetic structure of $SrFeO_3$ a set of 143 (28 unique) satellites was collected. The analysis confirmed the presence of a helical magnetic ordering of the $Fe^{4+}$-sublattice as found earlier.[1] Using the large number of absorption- and extinction-corrected magnetic structure factors, we were able to determine the magnetic moment with very good accuracy, as documented by the residual $R_F = 0.040$. The refined magnetic moment of the iron atoms was found to be $\mu_{exp} = 2.960(12)$ $\mu_B$ at 2 K. This value is larger than the value $\mu_{exp}$

= 2.7(4) $\mu_B$ obtained earlier at 77 K (Ref. 1), and it could be determined with improved precision. The total moment of the $Fe^{4+}$-ions ($3d^4$-electron configuration) is well below the free-ion value of 4.0 $\mu_B$. While this is not unexpected for a metallic compound, we note that the ordered moments of the $Fe^{3+}$-ions ($3d^5$-electron configuration) in the insulating charge-ordered ferrites $CaFeO_3$ and $Sr_4Fe_4O_{11}$ reach the values $\mu_{exp}$ = 3.5(1) $\mu_B$ and $\mu_{exp}$ = 3.55(5) $\mu_B$, respectively, [8,26] which are also well below the free-ion value of 5.0 $\mu_B$. Factors possibly contributing to the moment reduction are charge fluctuations in proximity to the metal-insulator transition, and the strong covalency between the Fe and O atoms in the $FeO_6$ and $FeO_5$ units.

Measurements as a function of temperature demonstrate that $\Delta$ decreases with increasing temperature (Fig. 3). This may at first sight appear surprising for a metallic system, where the helix propagation vector is expected to be tied to the dimensions of the Fermi surface. A recent study of $NbSe_2$ has, however, revealed that both the Fermi surface nesting vector and the charge density wave propagation vector depend substantially on temperature, and found a good match between both quantities.[28,29] A temperature dependent investigation of the Fermi surface of $SrFeO_3$ would therefore be an interesting subject of future work.

Figure 3 also shows that the temperature dependence of the incommensurability does not explain the difference between the saturated value $\Delta$ = 0.129 found in the present work and the value $\Delta$ = 0.112 at $T$ = 77 K reported earlier.[1] The difference probably reflects a different charge carrier concentration in the sample investigated earlier, possibly due to residual oxygen vacancies. We note, however, that $\Delta$ in the cubic volume fraction of our oxygen-deficient samples does not differ substantially from the one in the $SrFeO_{3.00}$ sample (Fig. 3, right panel). Rather, the presence of magnetic phase-IV reflections suggests that even a small number of residual oxygen defects segregates in our samples, while they may have been randomly incorporated in the earlier samples, possibly due to a different microstructure caused by a different heat treatment.

### 2. Phase II

The right part of Fig. 2 shows the neutron diffraction data of the single-crystalline sample $SrFeO_{2.87}$. The data were measured at the triple-axis spectrometer Rita II along the $[1,1,1]_{cub}$ direction around the nuclear Bragg reflection $(0, 0, 1)_{cub}$. As discussed in Section III.A above, this crystal consists of a phase mixture with predominantly tetragonal $Sr_8Fe_8O_{23}$ and cubic $SrFeO_3$ with a volume ratio of 74(2) : 26(2). This is indicated by the appearance of two sets of magnetic Bragg reflections with an incommensurate splitting of $\Delta = 0.13$ for the cubic phase, and a second set of Bragg reflections with $\Delta = 0.20$ that disappear upon heating above the charge ordering transition, $T_S = 75(2)$ K, of the tetragonal phase $Sr_8Fe_8O_{23}$ (Section III.B). As found for cubic $SrFeO_3$, the propagation vector of the magnetic "phase II" points along the body diagonal of the cubic unit cell. Only a total number of 7 magnetic reflections could be collected on the triple-axis spectrometer. Therefore we were not able to precisely determine the magnetic structure of this phase. However the refinements showed that the moments are aligned perpendicular to the propagation vector $k = (0.20, 0.20, 0.20)_{cub}$, indicating a similar helical structure as the one found in the cubic ferrite $SrFeO_3$.

For $T > T_S$, the phase-II satellites evolve into weak diffuse features indicative of correlated phase-II fluctuations in the tetragonal volume fraction of the $SrFeO_{2.87}$ crystal. Interestingly, the wave vector characterizing these fluctuations deviates from the commensurate value $\Delta = 0.20$ and is strongly temperature dependent. With increasing temperature, the diffuse features progressively approach and eventually merge with the Bragg peaks of the helical order in the cubic volume fraction of the crystal. This supports our conclusion that the $SrFeO_3$ and $Sr_8Fe_8O_{23}$ structures are segregated on a nanoscopic length scale, and that the magnetic ordering phenomena in both phases are influenced by exchange interactions across the phase boundaries.

### 3. Phase III

Our analysis of the data obtained on the powder of composition $SrFeO_{2.87}$ already discussed in Section III.B revealed a rather different magnetic ordering pattern in the $Sr_8Fe_8O_{23}$ volume fraction. Figure 6 shows a series of incommensurate magnetic Bragg reflections at the $2\theta$-positions 11.2°, 13.6°, 18.9° and 29.6°, which are present only for temperatures below the charge-ordering transition $T_S = 75$ K and can therefore be ascribed to the monoclinic structure of $Sr_8Fe_8O_{23}$. Using the propagation vector $k = (0.687, 0, 0.326)_{mon}$ these reflections could be indexed as $(0.687, 0, 0.326)_M$, $(0.313, 0, 0.674)_M$, $(1.313, 0, –0.326)_M$ and $(0.687, 2, 0.326)_M$. From our high-resolution neutron powder patterns collected at 11 K, we were able to determine the components of the propagation vector with standard deviations of $\sigma = \pm 0.001$. In the cubic notation, the propagation vector can be written as $(0.169, 0.169, 0.169)_{cub}$ and is therefore directed along the [111] direction of the cubic unit cell, as found above for phases I and II.

Since magnetic intensity only appears for magnetic reflection with $k = 2n$, antiferromagnetic coupling of the iron atoms with distances $d$(Fe-Fe) = ½$b$ along the monoclinic $b$-axis can be excluded. In order to determine the magnetic structure of this phase of $Sr_8Fe_8O_{23}$ we first tried the helical model of $SrFeO_3$. In this structure the magnetic moments of the iron atoms form ferromagnetic planes perpendicular to the propagation vector ***k***. In the monoclinic setting the real Fourier component is aligned within the $ac$-plane, subtending an angle of −35.3° with the $c$-axis. The imaginary Fourier component is set parallel to the monoclinic $b$-axis. With this model, the magnetic intensity of the first reflection (0.687, 0, 0.326)$_M$ could be well fitted. But in the 2$\theta$-range up to 45° the calculated magnetic intensities were found to be zero at the positions of the reflections (0.313, 0, 0.674)$_M$, (1.313, 0, 0.326)$_M$ and (0.687, 2, 0.326)$_M$, in disagreement with the experimental data. This suggests that the incommensurate spin structure of the magnetic phase III is different from the helical ordering observed in $SrFeO_3$ and in phase II of $Sr_8Fe_8O_{23}$. Further trials showed that magnetic intensities could be generated for all four reflections using a model in which the moments of Fe11 in (0, 0, 0.255) and Fe31 in (0, ½, 0.25) are coupled antiparallel to the moments of the Fe12 atoms in (0, 0, −0.255) and Fe32 in (0, ½, −0.25) (Fig. 10). Note that the atoms Fe11 and Fe12 show a square-pyramidal coordination (FeO$_5$). This is analogous to the magnetic structure of $Sr_4Fe_4O_{11}$, where the iron moments are also coupled antiparallel in the dimers of corner-shared FeO$_5$-square pyramids.[8] However, the refinements did not yield satisfactory agreement between the observed and calculated intensities of all the reflections. Further trials showed that the fit could be considerably improved by refinement of the phase factor $\varphi$ (given in fractions of 2π), which is only allowed to vary for independent sites of magnetically ordered atoms. But a representation analysis following Bertaut (Ref. 30) for the space group $I2/m$ showed that all the atoms Fe1, Fe21, Fe22 and Fe3 split into two orbits. The refined phase factor $\varphi$ between the two sets (Fe11, Fe31) and (Fe12, Fe32) resulted in a value close to zero indicating that the iron moments of both sets are aligned almost parallel to each other. For the Fe21-atoms we found the phase factors $\varphi$ = 0.201(11) [Fe21 atoms at 4$f$: (¼,¼,¼) and (¼,–¼,¼)] and $\varphi$ = 0.093(11) [Fe22 atoms at 4$e$: (¼,¼,–¼) and (¼,–¼,–¼)], which are slightly shifted from the ideal values $\varphi$ = 0.246 and $\varphi$ = 0.082 fixed for $SrFeO_3$. In this structure, the alignment of the Fe spins in the two-dimensional network of slightly tilted Fe2O$_6$-octahedra in $Sr_8Fe_8O_{23}$ is similar to that one within the network of undistorted FeO$_6$-octahedra in $SrFeO_3$, but the ordered moment $\mu_{exp}$ = 1.69(3) $\mu_B$ is considerably smaller. However, the refinement of the helical spin structure resulted in a relatively large residual $R_M$ = 0.168 (defined as $R_M = \sum|I_{obs} - I_{cal}|/\sum I_{obs}$). In particular, the calculated intensity of the reflection (0.686, 2, 0.326)$_M$ (observed at $2\theta = 29.6°$) was found to be much smaller than observed, whereas the fit of the other three reflections (given above) was satisfactory.

In order to improve the refinement, the real and imaginary Fourier components of the iron moments were allowed to vary independently. This gave a much better agreement between observed and calculated intensities, resulting in a residual $R_M = 0.087$. The refined imaginary component reached a value close to zero. This implies that the magnetic moments of $Sr_8Fe_8O_{23}$ in the powder sample show a "spin density wave" (SDW) state (Fig. 10) with an amplitude of 2.54(4) $\mu_B$, considerably smaller than the saturation moments $\mu_{exp} = 2.960(12)$ $\mu_B$ of the $Fe^{4+}$-ion in $SrFeO_3$ and 3.55(5) $\mu_B$ of the $Fe^{3+}$-ion in $Sr_4Fe_4O_{11}$ [Ref. 8].

## *4. Phase IV*

For the refinement of the magnetic structure of phase IV we used the tetragonal setting of $Sr_8Fe_8O_{23}$, where the propagation vector is $k = (0, 0, ½)$. Thus the magnetic structure requires a doubling of the $c$-axis of the tetragonal unit cell. The strongest magnetic intensity could be observed for the reflection $(0, 0, ½)_{tetr}$ at 9.5° (Fig. 6). This suggests a ferromagnetic ordering of the iron moments (all at $z = 0.25$ and 0.75 in the unit cell) within the $ab$-plane, and a spin sequence along the $c$-axis $+ +, - -, + \ldots$ . First the magnetic structure was refined using a model where the moments are aligned parallel to the $a$-axis. With this model the observed and calculated intensities of the reflection $(0, 0, ½)_M$ were found to be in good agreement, whereas the calculated intensity of the $(0, 0, 1½)_{tetr}$ (at $2\theta = 28.9°$) was found to be too low. But it could be seen that the magnetic reflections $(2, 0, ½)_M$ and $(0, 2, ½)_M$ are very close to the position of the reflection $(0, 0, 1½)_{tetr}$. Therefore we assumed a magnetic ordering where the second magnetic component was aligned parallel to the $c$-axis. Due to the fact that magnetic intensity was only observed for magnetic reflections with $h, k = 2n$ an antiferromagnetic spin sequence $+ -, + -, + \ldots$ along the $a$-axis is expected. Then the lattice parameter $a$ of the magnetic cell is only ½ of that of the tetragonal unit cell, but identical to the $c$-axis of orthorhombic structure of $Sr_4Fe_4O_{11}$. With this model, where the moments were allowed to vary in the $ac$-plane, we finally obtained a satisfactory refinement resulting in a residual $R_M = 0.067$. It was found that the moments are tilted with respect to the $c$-axis with an angle of 54(1)°, which is considerably larger than the value 35.3° found for $SrFeO_3$ and $Sr_8Fe_8O_{23}$ (Fig. 11). In the tetragonal setting one cannot distinguish whether the iron moments are aligned in the $ac$- or the $bc$-plane.

In order to estimate the content of the magnetic phase IV we assumed a magnetic moment of the iron atoms of $\mu_{exp} = 3.0$ $\mu_B$. The phase analysis of the powder sample with an oxygen deficiency of $\delta = 0.13$ showed that the powder contains 78% phase III, 14% phase I, and 8% phase IV. These fractions could be determined with a statistical accuracy of 2%, but are subject to a systematic error arising from the unknown ordered moment of phase IV. As we have seen in Section III.A, these magnetic phases are not necessarily fully congruent with oxygen-vacancy ordered structures. For the powder sample with $\delta = 0.19$ it was found that the sample additionally contains phase VII, which is characteristic of orthorhombic $Sr_4Fe_4O_{11}$, while phase I is absent. For this powder sample the phase analysis resulted in the following composition: 45% phase III, 40% phase VII, and 15% phase IV. As discussed in Section III.A, phase IV may belong to a new ferrate structure, which could not be identified in the present study. In mixed-phase samples, it is quite difficult to determine the crystal-structure of minority phases, because the strong reflections of all the different phases overlap, and the superstructure reflections of the minority phases are rather weak and difficult to measure with good accuracy.

## IV. CONCLUSIONS

Our experiments have revealed a surprising variety of magnetic ordering patterns in oxygen-deficient $SrFeO_{3-\delta}$. Table III lists the basis properties of the seven structures that could be identified, and Figure 11 depicts those structures for which a refinement could be obtained. This variety is a consequence of a confluence of several factors including competing magnetic interactions, charge fluctuations, and the ramified bond network created by the oxygen vacancy arrays. In order to understand the influence of these factors, it is useful to begin our discussion by considering $BiFeO_3$, an insulating compound with $Fe^{3+}$ ions that crystallizes in a rhombohedrally distorted, pseudocubic perovskite structure. $BiFeO_3$ exhibits nearly collinear antiferromagnetism with a high Néel temperature of $T_N = 643$ K.[31] (A long-wavelength cycloidal modulation is induced by exchange anisotropies due the rhombohedral lattice distortion.[32]) Its close analogue $NdFeO_3$ exhibits an even larger $T_N = 760$ K.[33] The high ordering temperatures reflect strong, unfrustrated superexchange interactions in a three-dimensional bond network. The sheets of trivalent Fe2-atoms in $Sr_4Fe_4O_{11}$ can be regarded as a two-dimensional segment of this network, cut along the [110] planes of the cubic cell. As a result, Fe2-spins along the orthorhombic $c$-axis are coupled via 180° bonds like those in (Bi,Nd)FeO$_3$ while those along the *a*-axis are coupled via weaker 90° bonds. The coupling between the antiferromagnetic Fe2-atom layers via the intervening Fe1-atom chains is presumably even weaker. This dilution of the exchange bond network is responsible for the much lower ordering temperature $T_N$ = 232 K of Sr$_4$Fe$_4$O$_{11}$ compared to BiFeO$_3$ and NdFeO$_3$. The resulting amplitude modulation of the ordered moment, which implies paramagnetic Fe1-atom chains even as $T \to 0$, deserves further investigation.

The incommensurate helical order of SrFeO$_3$, with uniform Fe$^{4+}$ valence, can be traced to a delicate competition between ferromagnetic double-exchange and antiferromagnetic super-exchange interactions on the verge of a metal-insulator transition.[34] In fact, its close analogue CaFeO$_3$, whose electronic bandwidth is lowered by an orthorhombic tilt distortion of the FeO$_6$ octahedra, undergoes a metal-insulator transition associated with charge-disproportionation at 290 K, and a transition to helical magnetism at $T_N$ = 115 K.[25,35-37] The helix propagation vector of ***k*** = (0.161, 0.161, 0.161)$_{cub}$ in the insulating state of CaFeO$_3$ reflects a smaller ratio of double-exchange and super-exchange interactions compared to the one in the metallic state of SrFeO$_3$, where ***k*** = (0.128, 0.128, 0.128)$_{cub}$.

In the range 0 < $\delta$ < 0.25 both frustration mechanisms are present and conspire to suppress the magnetic ordering temperatures even further, into the range 60 K ≤ $T$ ≤ 75 K. The low-temperature ordered moments follow an analogous trend. Among the resulting magnetic phases II-VI, only phases II and III could be uniquely associated with a vacancy-ordered structure, namely Sr$_8$Fe$_8$O$_{23}$. In this structure, the vacancy order induces segregation of the Fe valence into different lattice planes. In contrast to Sr$_4$Fe$_4$O$_{11}$, however, the network of exchange bonds between tetravalent Fe1- and Fe3-atoms is not as strongly disrupted as in Sr$_4$Fe$_4$O$_{11}$, and the sheets of Fe2-atoms are in the intermediate valence state Fe$^{3.5+}$ rather than Fe$^{3+}$. As a result, the magnetic ordering patterns in the monoclinic, insulating state for $T \leq T_S$ = 75 K resemble the helix states with propagation vector along the cubic [111] axis observed in SrFeO$_3$ and CaFeO$_3$. The incommensurability of phase III, $\Delta$ = 0.169, is larger than the one of SrFeO$_3$ and close the one of CaFeO$_3$, in accord with the expectation that the strength of the antiferromagnetic super-exchange increases relative to the ferromagnetic double-exchange in compounds with more localized electrons. In contrast to SrFeO$_3$ and CaFeO$_3$, however, phase III is a collinear, amplitude-modulated state. This aspect is qualitatively analogous to the antiferromagnetic state observed in Sr$_4$Fe$_4$O$_{11}$.

While phase III was observed in a SrFeO$_{2.87}$ powder sample with Sr$_8$Fe$_8$O$_{23}$ as majority vacancy-ordered phase, we found a different magnetic structure, phase II, in the Sr$_8$Fe$_8$O$_{23}$ volume fraction of a single crystal with the same oxygen stoichiometry. Phase II is a commensurate helix state with $\Delta = 0.2$. This indicates a subtle balance between amplitude-modulated and helix states in this range of oxygen deficiencies. Details of the oxygen distribution (such as the density of possible randomly incorporated oxygen defects in $Sr_8Fe_8O_{23}$) or the microstructure of the lattice appear to determine the balance between these nearly degenerate magnetic states.

Finally, we point out that the magnetic microstructure resulting from this delicate phase balance appears to exert a key influence on the magneto-transport properties. A prime example is the commensurate phase IV, which occupies only a very small volume fraction in nominally stoichiometric $SrFeO_{3.00}$ single crystals, but nonetheless triggers a large-scale modification of the domain structure of the incommensurate majority phase I. This, in turn, induces a resistivity jump $T = 60$ K, the onset of magnetic order in phase IV. The magnetic field dependence of the phase-IV ordering temperature is therefore presumably responsible for the large magnetoresistance in $SrFeO_{3.00}$ crystals.[11,12]

In the future, it will be interesting to explore the influence of chemical substitution (such as La for Sr and/or Co for Fe) pressure, and magnetic field on the interplay between oxygen vacancy order, magnetic microstructure, and magnetoresistance of $SrFeO_3$ and other ferrates.[15,38] We also note that a delicate dependence of the magnetic texture on external parameters and associated transport anomalies have been observed in other incommensurately modulated magnets such as $BiFeO_3$, MnSi, and MnGe.[17-19] A detailed understanding of these phenomena is an interesting target for future research.

## ACKNOWLEDGMENTS

We acknowledge fruitful discussions with T. Arima, D. Efremov, G. Khaliullin, D. Peets, and Y. Tokura, and thank N. Cavadini and D. Sheptyakov for help with the measurements at the Paul Scherrer Institut. We thank the German Science Foundation (DFG) for financial support under collaborative grant No. SFB/TRR 80. This paper is partly based on the results of the experiments carried out at the Swiss spallation neutron source SINQ, Paul Scherrer Institut, Villigen, Switzerland.

TABLE I. Results of the crystal structure refinements of $Sr_8Fe_8O_{23}$ from neutron powder diffraction ($\lambda = 1.158$ Å). The refinements of the data sets collected at 90 and 199 K were carried out in the tetragonal space group *I*4/*mmm*. For the data collected at 2 K the structure was refined in the monoclinic space group *I*2/*m* (*I* 1 2/*m* 1). In the last refinement the *z*-parameter of Fe3 was fixed at the value 0.25. Further, the parameters *x* and *y* of O3 as well as the thermal parameters $B_{is}$ of each element Sr, Fe, and O were constrained to be equal.

| | $Sr_8Fe_8O_{23}$ at 2 K, in *I*2/*m* | | | | $Sr_8Fe_8O_{23}$ at 90 K, in *I*4/*mmm* | | | | $Sr_8Fe_8O_{23}$ at 199 K, in *I*4/*mmm* | | | |
|---|---|---|---|---|---|---|---|---|---|---|---|---|
| | *x* | *y* | *z* | $B_{is}$ | *x* | *y* | *z* | $B_{is}$ | *x* | *y* | *z* | $B_{is}$ |
| Sr1(1) | 0.2568(14) | 0 | 0.9968(19) | 0.07(2) | 0.2580(4) | 0 | 0 | 0.14(1) | 0.2586(4) | 0 | 0 | 0.25(1) |
| Sr1(2) | 0 | 0.2606(15) | 0 | 0.07 | - | - | - | 0.14 | - | - | - | 0.25 |
| Sr2(1) | 0.2462(13) | 0 | 0.5037(20) | 0.07 | 0.2519(4) | 0 | ½ | 0.14 | 0.2536(3) | 0 | ½ | 0.25 |
| Sr2(2) | 0 | 0.2532(12) | ½ | 0.07 | - | - | - | 0.14 | - | - | - | 0.25 |
| Fe1 | 0.9995(12) | 0 | 0.2549(6) | 0.03(1) | 0 | 0 | 0.2546(6) | 0.05(1) | 0 | 0 | 0.2543(6) | 0.10(1) |
| Fe2(1) | ¼ | ¼ | ¼ | 0.03 | ¼ | ¼ | ¼ | 0.05 | ¼ | ¼ | ¼ | 0.10 |
| Fe2(2) | ¼ | ¼ | ¾ | 0.03 | - | - | - | - | - | - | - | - |
| Fe3 | 0.5005(13) | 0 | 0.250 | 0.03 | ½ | 0 | ¼ | 0.05 | ½ | 0 | ¼ | 0.10 |
| O1 | 0 | 0 | ½ | 0.25(2) | 0 | 0 | ½ | 0.30(1) | 0 | 0 | ½ | 0.42(1) |
| O2(1) | 0.1212(10) | 0.1176(12) | 0.2127(8) | 0.25 | 0.1201(3) | 0.1201 | 0.2227(4) | 0.30 | 0.1203(3) | 0.1203 | 0.2228(4) | 0.42 |
| O2(2) | 0.1173(11) | 0.1221(11) | 0.7618(10) | 0.25 | - | - | - | 0.30 | - | - | - | 0.42 |
| O3 | 0.2393(4) | 0.2393 | 0.5027(21) | 0.25 | 0.2398(4) | 0.2398 | ½ | 0.30 | 0.2398(4) | 0.2398 | ½ | 0.42 |
| O4(1) | 0.1261(12) | 0.6210(17) | 0.2496(19) | 0.25 | 0.1238(5) | 0.6238 | ¼ | 0.30 | 0.1237(4) | 0.6237 | ¼ | 0.42 |
| O4(2) | 0.1293(12) | 0.6204(17) | 0.7439(16) | 0.25 | - | - | - | 0.30 | - | - | - | 0.42 |
| O5(1) | ½ | 0 | 0 | 0.25 | ½ | 0 | 0 | 0.30 | ½ | 0 | 0 | 0.42 |
| O5(2) | 0 | ½ | 0 | 0.25 | - | - | - | 0.30 | - | - | - | 0.42 |

TABLE II. Interatomic distances (in Å) between the oxygen and iron atoms in tetragonal and monoclinic $Sr_8Fe_8O_{23}$. The Fe2- and Fe3-atoms show an octahedral coordination ($FeO_6$). Due the oxygen deficiency the Fe1-atom shows a square pyramidal coordination ($FeO_5$). The lattice constants, cell volumes, and bond angles of $Sr_8Fe_8O_{23}$ are also given. The values determined at 2, 90, and 199 K are compared with the room temperature values presented by Hodges et al.[6]

| $Sr_8Fe_8O_{23}$ | at 2 K ($I2/m$) | at 90 K ($I4/mmm$) | at 199 K ($I4/mmm$) | at RT ($I4/mmm$) |
|---|---|---|---|---|
| $d_{Fe1-O21}$ | 1.873(9) ×2 | 1.868(3) ×4 | 1.871(3) ×4 | 1.851(4) ×4 |
| $d_{Fe1-O22}$ | 1.847(9) ×2 | 1.868 | 1.871 | 1.851 |
| $d_{Fe1-O1}$ | 1.882(5) ×1 | 1.886(4) ×1 | 1.889(3) ×1 | 1.926(1) ×1 |
| $d_{O1-O21}$ | 2.874(8) ×2 | 2.823(3) ×4 | 2.826(3) ×4 | 2.820(3) ×4 |
| $d_{O1-O22}$ | 2.729(8) ×2 | 2.823 | 2.826 | 2.820 |
| $d_{O21-O21}$ | 2.564(18) ×1 | 2.619(4) ×4 | 2.624(4) ×4 | 2.601(4) ×4 |
| $d_{O22-O22}$ | 2.663(18) ×1 | 2.619 | 2.624 | 2.601 |
| $d_{O21-O22}$ | 2.608(16) ×2 | 2.619 | 2.624 | 2.601 |
| | | | | |
| $d_{Fe21-O21}$ | 2.034(9) ×2 | 2.013(3) ×2 | 2.013(3) ×2 | 2.036(3) ×2 |
| $d_{Fe21-O41}$ | 1.950(9) ×2 | 1.945(5) ×2 | 1.948(6) ×2 | 1.952(3) ×2 |
| $d_{Fe21-O3}$ | 1.947(16) ×2 | 1.9274(3) ×2 | 1.9292(4) ×2 | 1.931(1) ×2 |
| $d_{O21-O3}$ | 2.737(13) ×2 | 2.755(4) ×2 | 2.757(5) ×2 | 2.772(5) ×2 |
| $d_{O21-O3}$ | 2.893(14) ×2 | 2.819(4) ×2 | 2.819(4) ×2 | 2.839(4) ×2 |
| $d_{O41-O3}$ | 2.753(15) ×2 | 2.738(5) ×2 | 2.742(5) ×2 | 2.745(2) ×4 |
| $d_{O41-O3}$ | 2.759(15) ×2 | 2.738(5) ×2 | 2.742 ×2 | 2.745 |
| $d_{O21-O41}$ | 2.770(16) ×2 | 2.799(6) ×2 | 2.801(6) ×2 | 2.820(3) ×4 |
| $d_{O21-O41}$ | 2.865(9) ×2 | 2.799 ×2 | 2.801 ×2 | 2.820 |
| | | | | |
| $d_{Fe22-O22}$ | 2.011(9) ×2 | 2.013(3) ×2 | 2.013(3) ×2 | 2.036(3) ×2 |
| $d_{Fe22-O42}$ | 1.931(9) ×2 | 1.945(5) ×2 | 1.948(6) ×2 | 1.952(3) ×2 |
| $d_{Fe22-O3}$ | 1.906(16) ×2 | 1.9274(3) ×2 | 1.9292(4) ×2 | 1.931(1) ×2 |
| $d_{O22-O3}$ | 2.714(14) ×2 | 2.755(4) ×2 | 2.757(5) ×2 | 2.772(5) ×2 |
| $d_{O22-O3}$ | 2.826(13) ×2 | 2.819(4) ×2 | 2.819(4) ×2 | 2.839(4) ×2 |
| $d_{O42-O3}$ | 2.686(15) ×2 | 2.738(5) ×2 | 2.742(5) ×2 | 2.745(2) ×4 |
| $d_{O42-O3}$ | 2.741(16) ×2 | 2.838(4) ×2 | 2.742 ×2 | 2.745 |
| $d_{O22-O42}$ | 2.762(14) ×2 | 2.799(6) ×2 | 2.801(6) ×2 | 2.820(3) ×4 |
| $d_{O22-O42}$ | 2.813(13) ×2 | 2.799 ×2 | 2.801 ×2 | 2.820 |
| | | | | |
| $d_{Fe3-O41}$ | 1.909(9) ×2 | 1.909(5) ×4 | 1.909(5) ×4 | 1.912(3) ×4 |
| $d_{Fe3-O42}$ | 1.923(10) ×2 | 1.909 | 1.909 | 1.912 |
| $d_{Fe3-O5}$ | 1.920(1) ×2 | 1.9211(3) ×2 | 1.9227(3) ×2 | 1.925(1) ×2 |
| $d_{O41-O51}$ | 2.708(10) ×2 | 2.708(4) ×8 | 2.710(4) ×8 | 2.713(2) ×8 |
| $d_{O41-O52}$ | 2.701(10) ×2 | 2.708 | 2.710 | 2.713 |
| $d_{O42-O51}$ | 2.685(10) ×2 | 2.708 | 2.710 | 2.713 |
| $d_{O42-O52}$ | 2.755(10) ×2 | 2.708 | 2.710 | 2.713 |
| $d_{O41-O41}$ | 2.638(8) ×1 | 2.699(8) ×4 | 2.700(8) ×4 | 2.704(4) ×4 |
| $d_{O42-O42}$ | 2.626(8) ×1 | 2.699 | 2.700 | 2.704 |
| $d_{O41-O42}$ | 2.785(17) ×2 | 2.699 | 2.700 | 2.704 |
| | | | | |
| $d_{av(Fe1-O)}$ | 1.864(8) | 1.871(3) | 1.875(4) | 1.866(2) |
| $d_{av(Fe21-O)}$ | 1.977(11) | 1.962(3) | 1.963(3) | 1.976(2) |
| $d_{av(Fe22-O)}$ | 1.949(11) | 1.962 | 1.963 | 1.976 |

| $d_{av(Fe3-O)}$ | 1.917(7) | 1.913(4) | 1.914(4) | 1.916(2) |
|---|---|---|---|---|
| $a$ [Å] | 10.8995(6) | 10.90005(10) | 10.91050(12) | 10.929(1) |
| $b$ [Å] | 10.9033(6) | 10.90005 | 10.91050 | 10.929 |
| $c$ [Å] | 7.67873(14) | 7.68424(14) | 7.69092(16) | 7.698(1) |
| $\beta$ [°] | 90.0764(25) | 90 | 90 | 90 |
| $V$ [Å$^3$] | 912.54(8) | 912.97(2) | 915.52(3) | 919.47(21) |

TABLE III. Magnetic phases in the system SrFeO$_{3-\delta}$ ($0 \leq \delta \leq 0.25$) found by single-crystal or powder neutron diffraction, as indicated. In this system the magnetic moments of the iron atoms show commensurate antiferromagnetic (AF), helical, and amplitude-modulated spin density wave (SDW) structures. Note that not all the observed spin order types could be attributed to the known crystal structures.

| Magnetic phase | Propagation vector $k$ | Type of spin order | $T_N$ (K) | Samples | Crystal structure |
|---|---|---|---|---|---|
| Phase I | (0.13, 0.13, 0.13)$_{cub}$ | helical | 133(1), 134 (Ref. 1) | SrFeO$_{2.87-3.00}$ (crystal) | SrFeO$_3$ |
| Phase II | (0.20, 0.20, 0.20)$_{cub}$ | helical | 75(2) | SrFeO$_{2.87}$ (crystal) | Sr$_8$Fe$_8$O$_{23}$ |
| Phase III | (0.17, 0.17, 0.17)$_{cub}$ (0.687, 0, 0.326)$_{mon}$ | SDW | 75(2) | SrFeO$_{2.87}$ (powder) | Sr$_8$Fe$_8$O$_{23}$ |
| Phase IV | (0, 0, ¼)$_{cub}$ (0, 0, ½)$_{tetr}$ | canted AF | 65(4) | SrFeO$_{2.77-3.00}$ (crystal) | unknown |
| Phase V | (0.30, 0.30, 0.75)$_{cub}$ | SDW or helical | 110(4) | SrFeO$_{2.77}$ (crystal) | unknown |
| Phase VI | (0.79, 0.79, 0)$_{cub}$ | SDW or helical | 60(5) | SrFeO$_{2.77}$ (crystal) | unknown |
| Phase VII | 0 | AF | 232(2), 232(4) (Ref. 8) | SrFeO$_{2.77}$ (crystal) | Sr$_4$Fe$_4$O$_{11}$ |

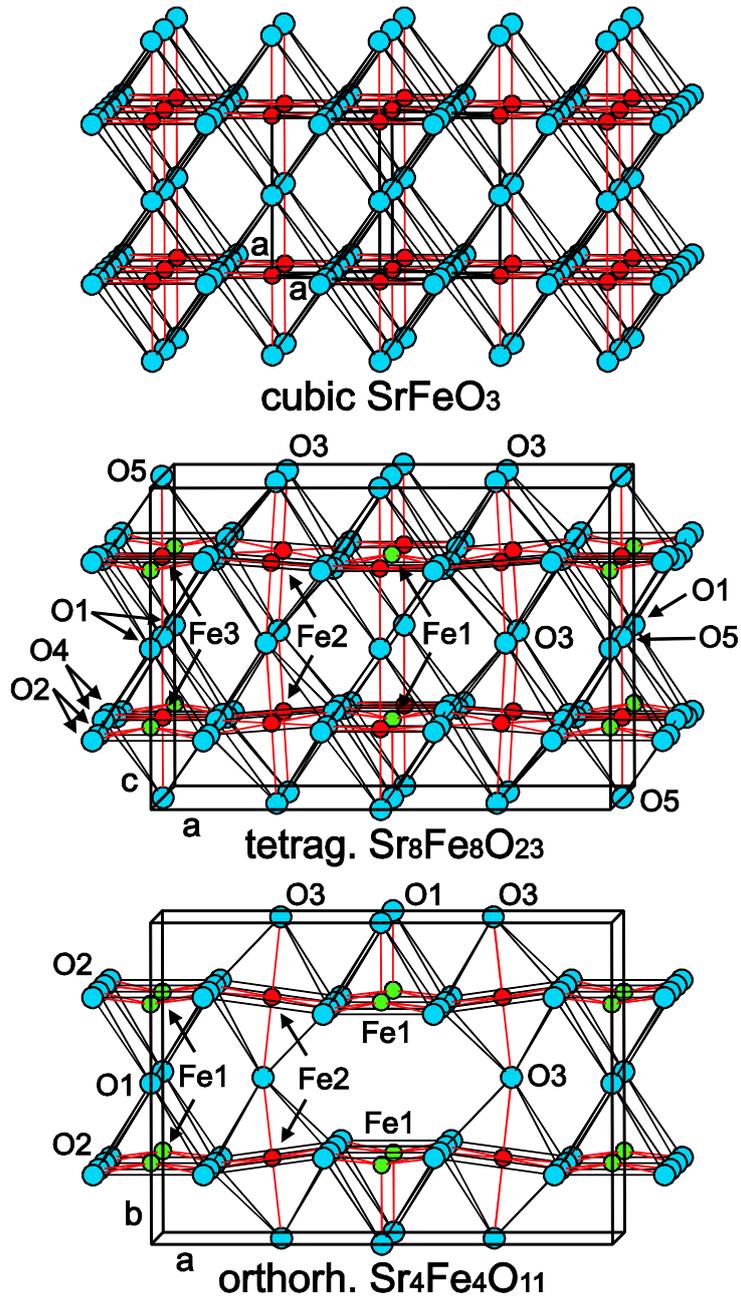

FIG. 1 (Color online). Crystal structures of vacancy ordered ferrates in the system SrFeO$_{3-x}$. For a clearer presentation, the Sr$^{2+}$-ions are not shown. The network of cubic SrFeO$_3$ is formed by undistorted corner-sharing FeO$_6$-octahedra. In tetragonal (*I*-centered) Sr$_8$Fe$_8$O$_{23}$, every second O-position along [1,0,0] and [0,1,0] (starting from the origin and the lattice

point ½,½,½) is unoccupied, so that sequences of alternating $FeO_6$ octahedra and $FeO_5$ square-pyramids are generated. (Fe-atoms are marked in red and green). A further oxygen deficiency within these chains is found in orthorhombic $Sr_4Fe_4O_{11}$, where one finds only corner shared dimers of $FeO_5$-square-pyramids along [001].

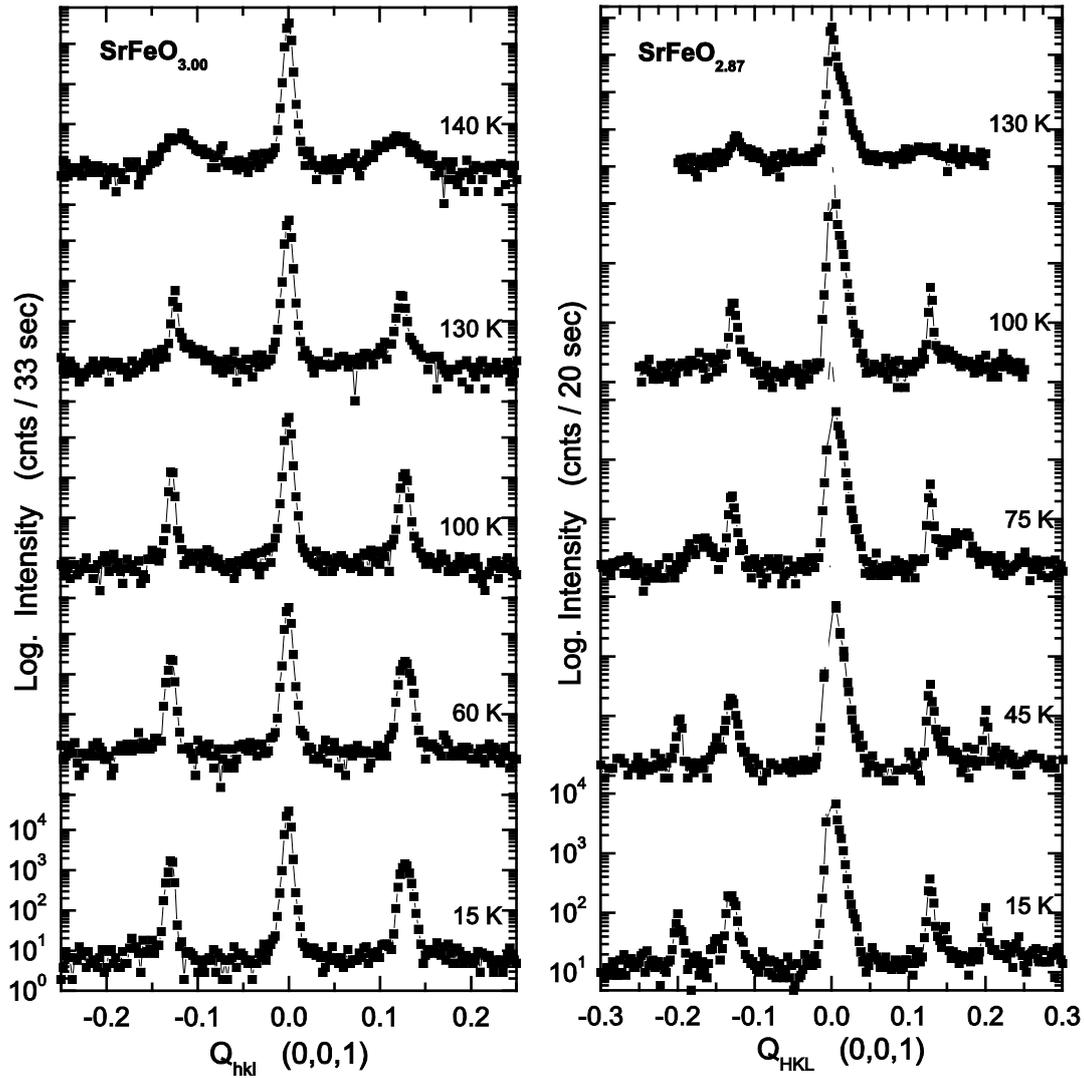

FIG 2. Neutron diffraction data on the single crystals of composition $SrFeO_{3.00}$ and $SrFeO_{2.87}$ measured at various temperatures. The scans were performed along the [1,1,1] direction around the structural Bragg reflection $(0, 0, 1)_{cub}$. With decreasing temperature magnetic reflections of phase I appeared in both samples. In the oxygen deficient crystal $SrFeO_{2.87}$ an additional set of magnetic reflection of phase II could be observed.

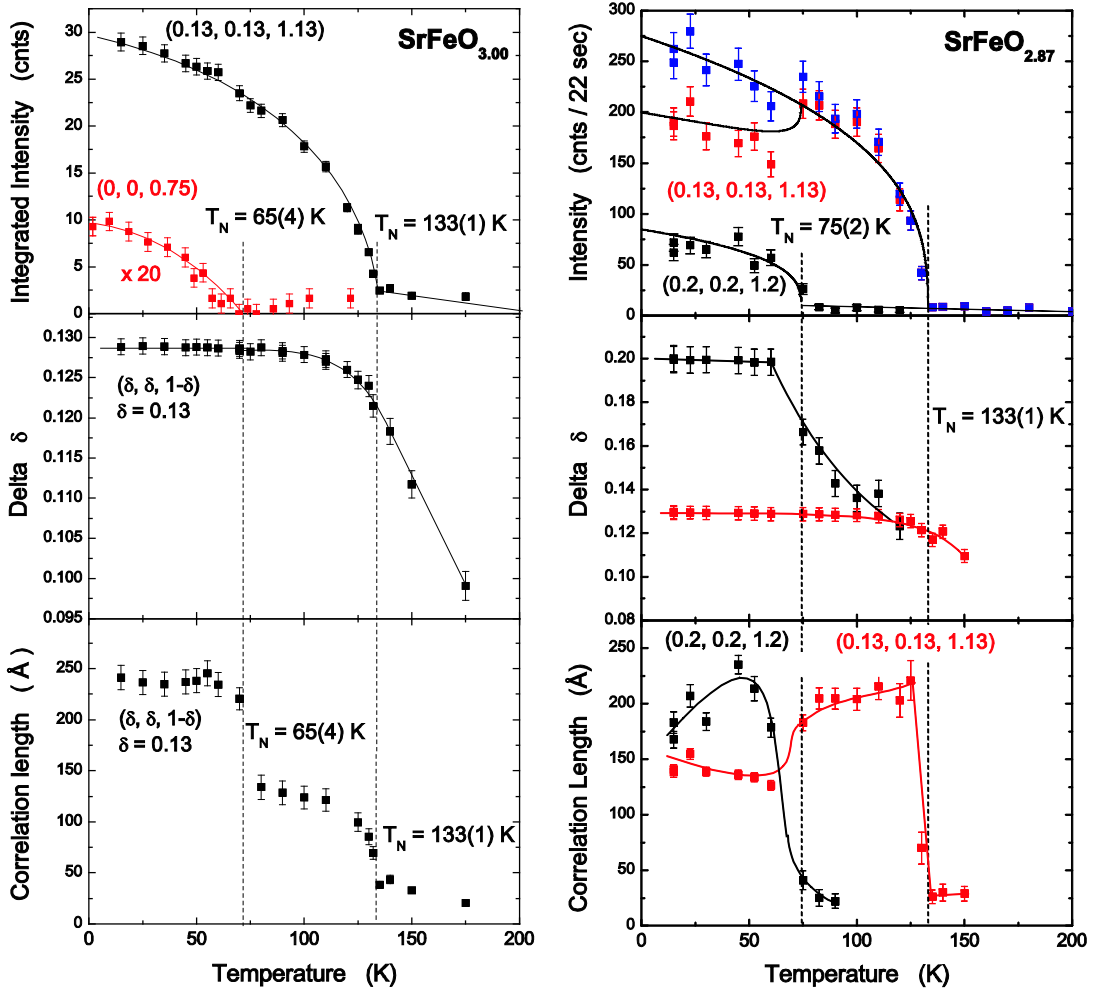

FIG. 3 (Color online). Temperature dependence of the intensity, incommensurability, and magnetic correlation length of the helical magnetic Bragg reflections of the magnetic phase I in the single crystal $SrFeO_{3.00}$ (left) and the magnetic phases I and II in $SrFeO_{2.87}$ (right). For the sample $SrFeO_{3.00}$ very weak intensity was found on the position $(0, 0, ¾)_{cub}$ indicating the presence of an additional magnetic phase (denoted as phase IV in the text) for temperatures below 65(4) K. The solid lines in the upper panels are the results of power-law fits to the measured intensity. It is interesting to note that the incommensurability is temperature dependent close to the magnetic phase transition.

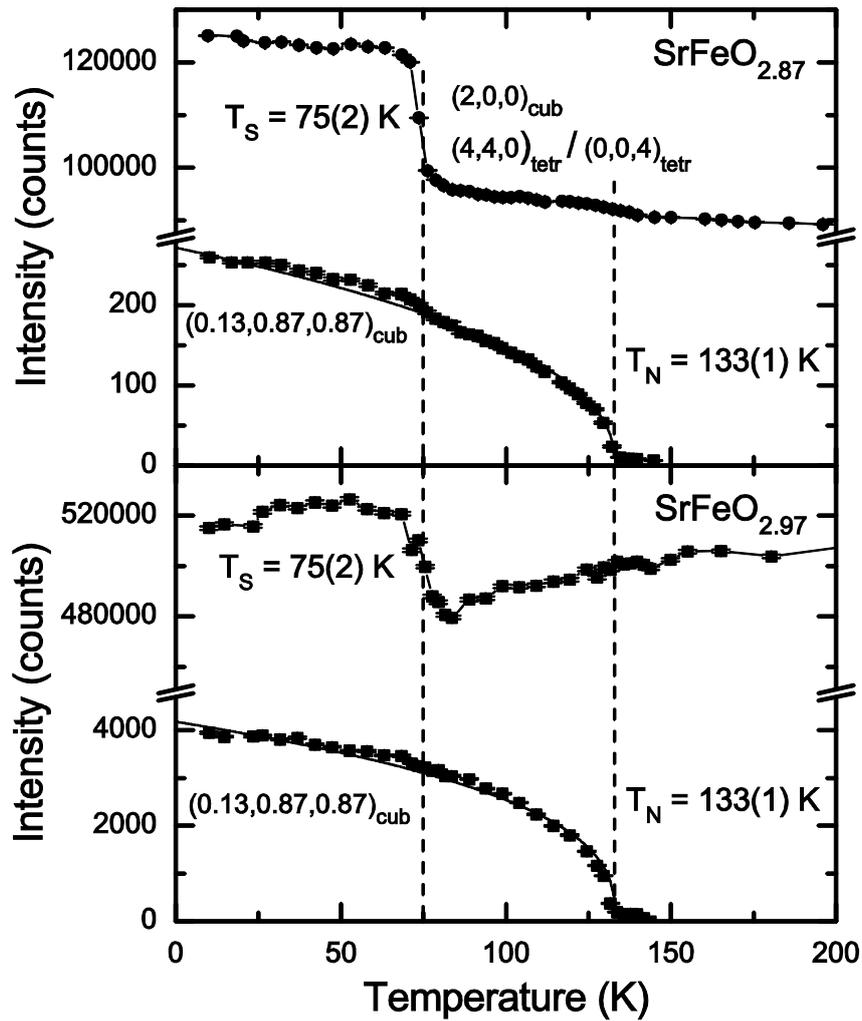

FIG. 4 . Temperature dependence of the strong nuclear reflection $(200)_{cub}$ of SrFeO$_{3-\delta}$ and the magnetic reflection $(0.13, 0.87, 0.87)_{cub}$ of the cubic phase SrFeO$_3$ (denoted as phase I). For the samples with lower oxygen content, the reflection $(2, 0, 0)_{cub}$ [$(4, 4, 0)_{tetr}$/$(0, 0, 4)_{tetr}$] shows a strong change of intensity at $T_S = 75(2)$ K indicating the presence of a structural phase transition. This transition is more pronounced in the sample SrFeO$_{2.87}$, suggesting that this transition occurs in the tetragonal phase Sr$_8$Fe$_8$O$_{23}$.

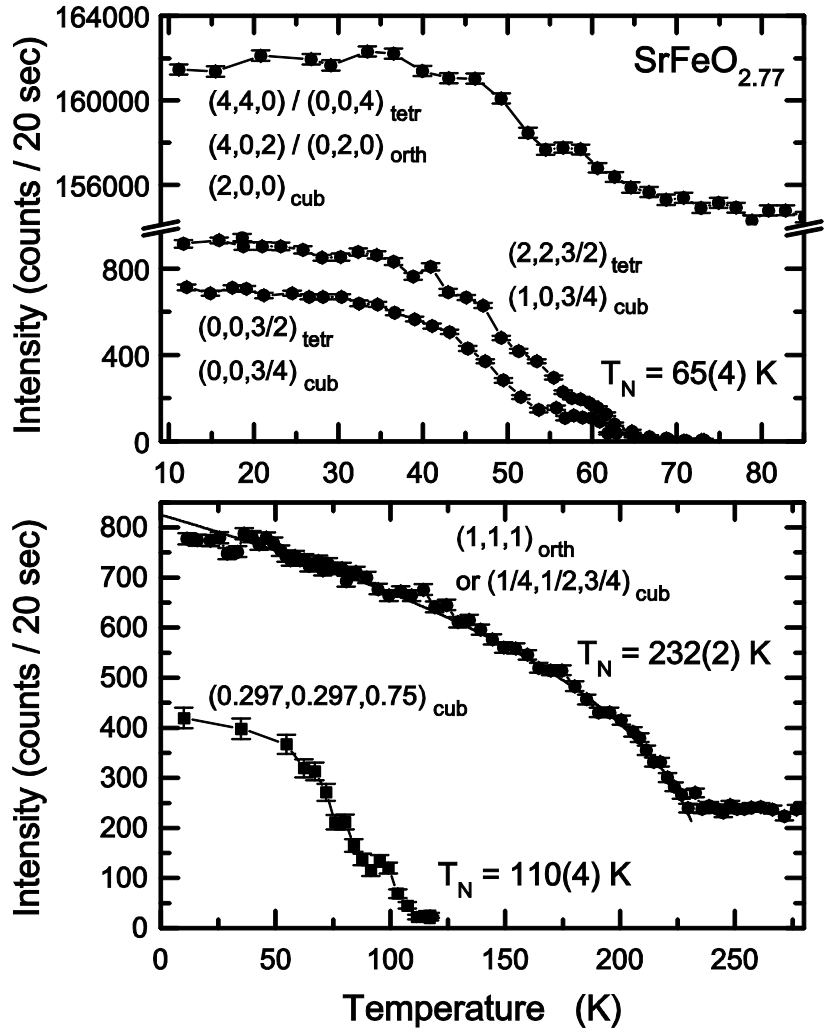

FIG. 5. Temperature dependence of the reflections $(2, 0, 0)_{cub}$, $(0, 0, ¾)_{cub}$, $(1, 0, ¾)_{cub}$, and $(¼, ½, ¾)_{cub}$ of $SrFeO_{2.77}$. Magnetic intensity of the $(¼, ½, ¾)_{cub}$ [$(1, 1, 1)_{orth}$ in the orthorhombic setting] appears spontaneously at 232(2) K due to the onset of antiferromagnetic ordering of the Fe-atoms (phase VII) in orthorhombic $Sr_4Fe_4O_{11}$. A second magnetic transition was observed at $T_N$ = 65(3) K. This can be ascribed to an antiferromagnetic ordering in a second phase, which could not be characterized so far. Furthermore, magnetic Bragg reflections with $T_N$ = 110(4) K were observed at the incommensurate positions $(h \pm 0.297, k \pm 0.297, \ell \pm 0.25)_{cub}$ (phase V).

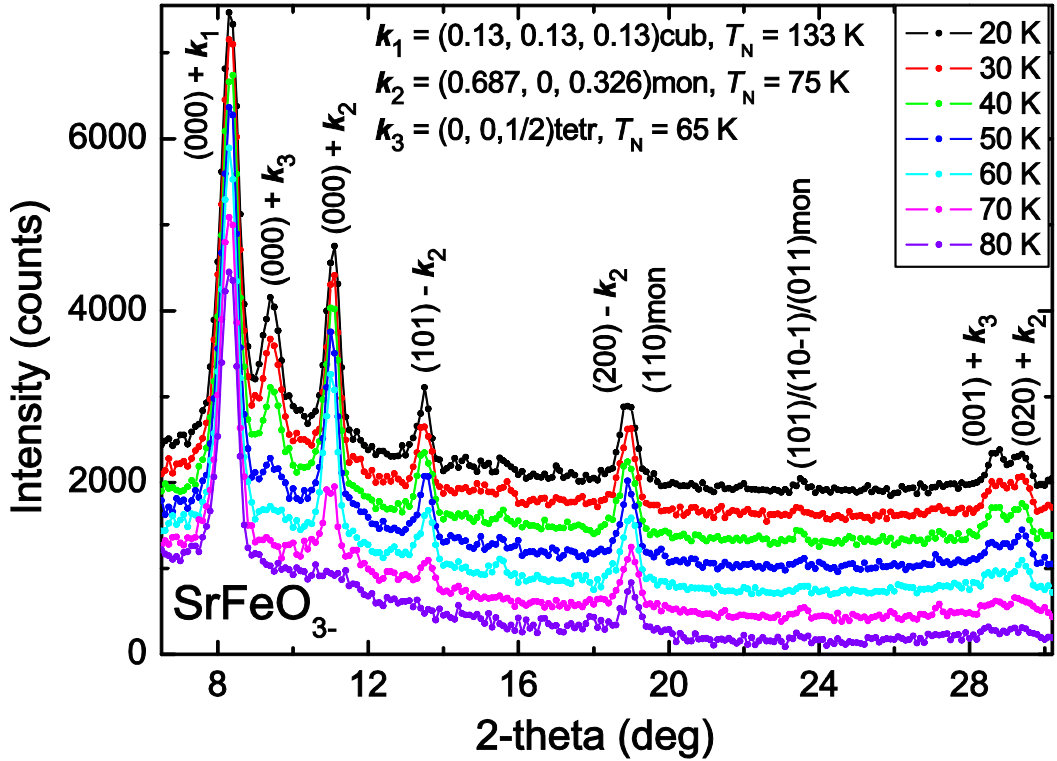

FIG. 6 (Color online). Neutron powder patterns of $SrFeO_{3-\delta}$ with $\delta$ = 0.13 at different temperatures between 20 and 80 K. The traces for temperatures below 80 K have been shifted for clarity. The neutron wavelength was λ = 2.58 Å. Below $T_N$ = 133(1) K magnetic intensity appears at the position $(0.13, 0.13, 0.13)_{cub}$ due to the incommensurate ordering (phase I) of the end member $SrFeO_3$. Below $T_N$ = 75(2) K incommensurate magnetic order of the iron moments (phase III) with $k_2$ = $(0.687, 0, 0.326)_{mon}$ (using the monoclinic unit cell of low-temperature structure of $Sr_8Fe_8O_{23}$). Further commensurate magnetic reflections are observable below $T_N$ = 65(4) K, suggesting the presence of a third magnetic phase (phase IV) with $k_3$ = $(0, 0, ½)_{tetr}$ of another oxygen deficient ferrate, which could not be characterized so far.

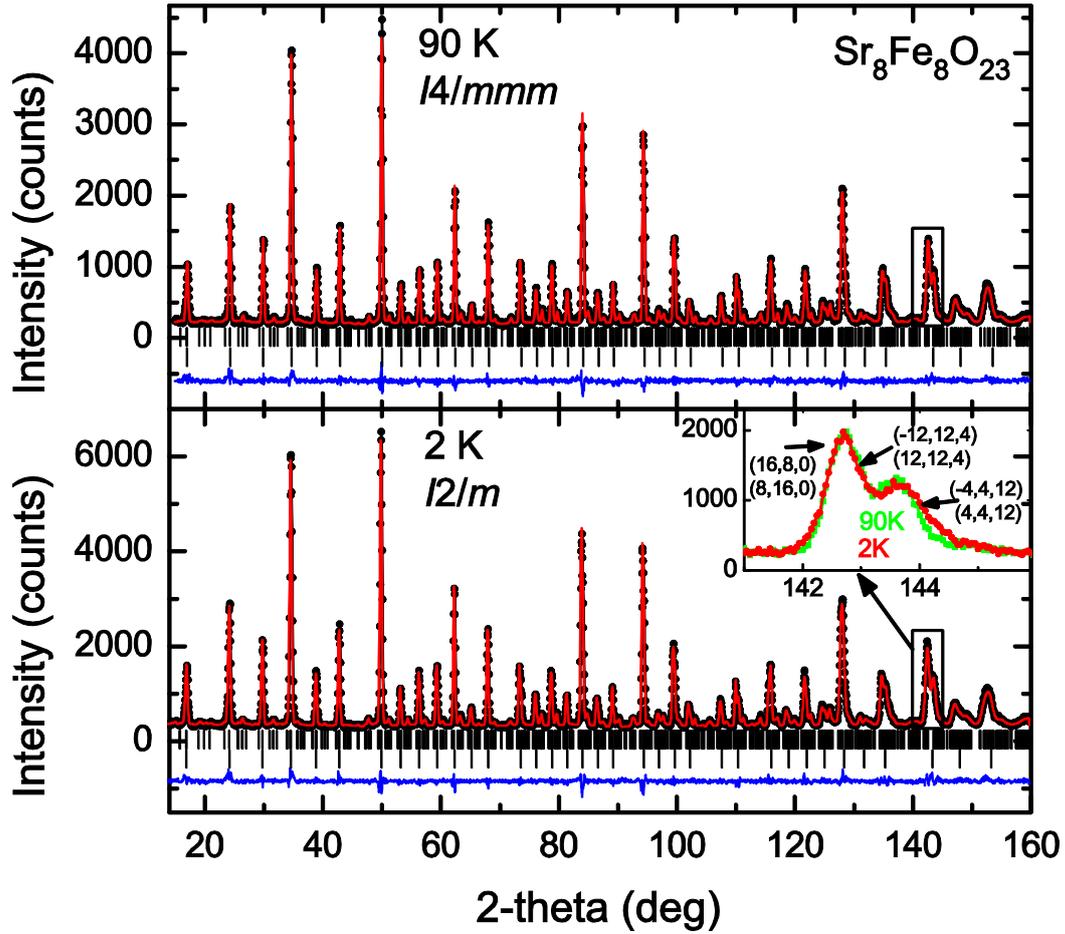

FIG. 7 (Color online). Neutron powder patterns of $Sr_8Fe_8O_{23}$. From the data sets collected at 2 and 90 K the crystal structure was refined in the monoclinic and tetragonal space groups $I2/m$ and $I4/mmm$, respectively. The calculated patterns (red) are compared with the observed ones (black circles). The difference patterns (blue) as well as the peak positions (black bars) of $Sr_8Fe_8O_{23}$ (above) and the cubic phase $SrFeO_3$ (below) are given in the lower part of each diagram. The inset indicates that the $2\theta$-positions of reflections ($h, k, \ell$) with large $\ell$ showed the strongest changes below the structural phase transition.

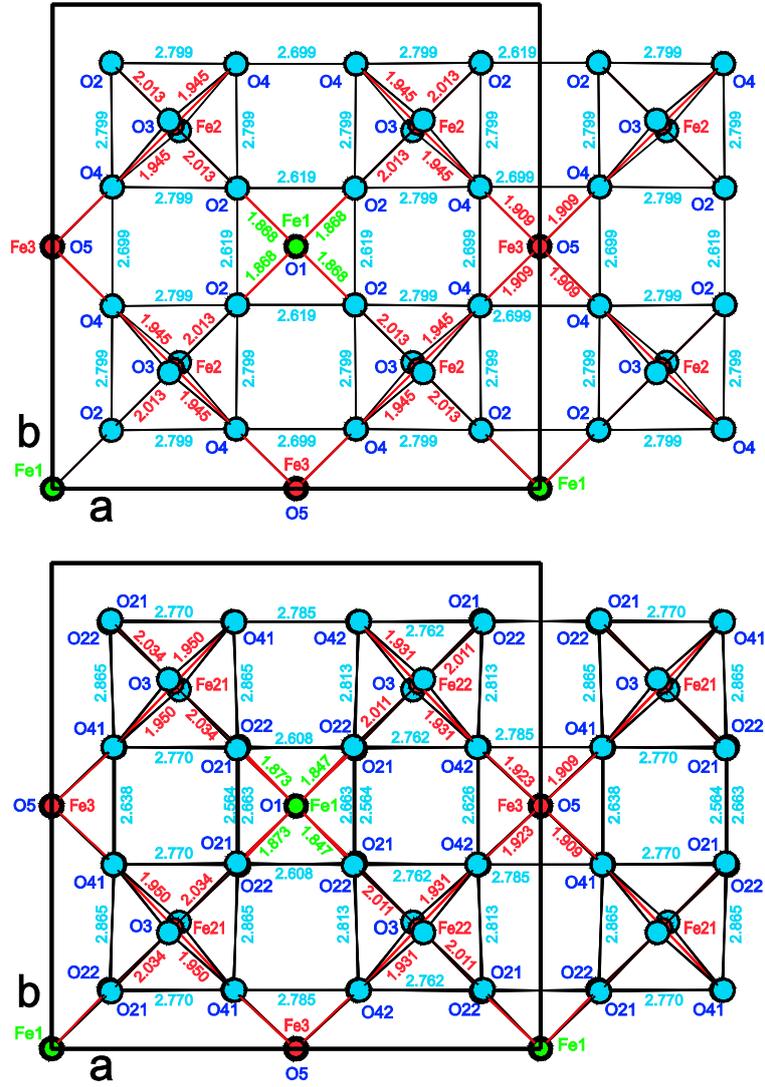

FIG. 8 (Color online). Projection of the high-temperature tetragonal (top) and the low-temperature monoclinic (bottom) crystal structure of $Sr_8Fe_8O_{23}$ along [001]. The network of distorted $FeO_6$-octahedra in the $ab$-plane is shown at the level $z = 0.25$. In order to see the displacements of the O2-atoms in the monoclinic structure, the O2-atoms of the neighboring $FeO_6$-octahedra at $z = 0.75$ are also shown. The values of particular interatomic Fe-O and O-O bond distances are also given.

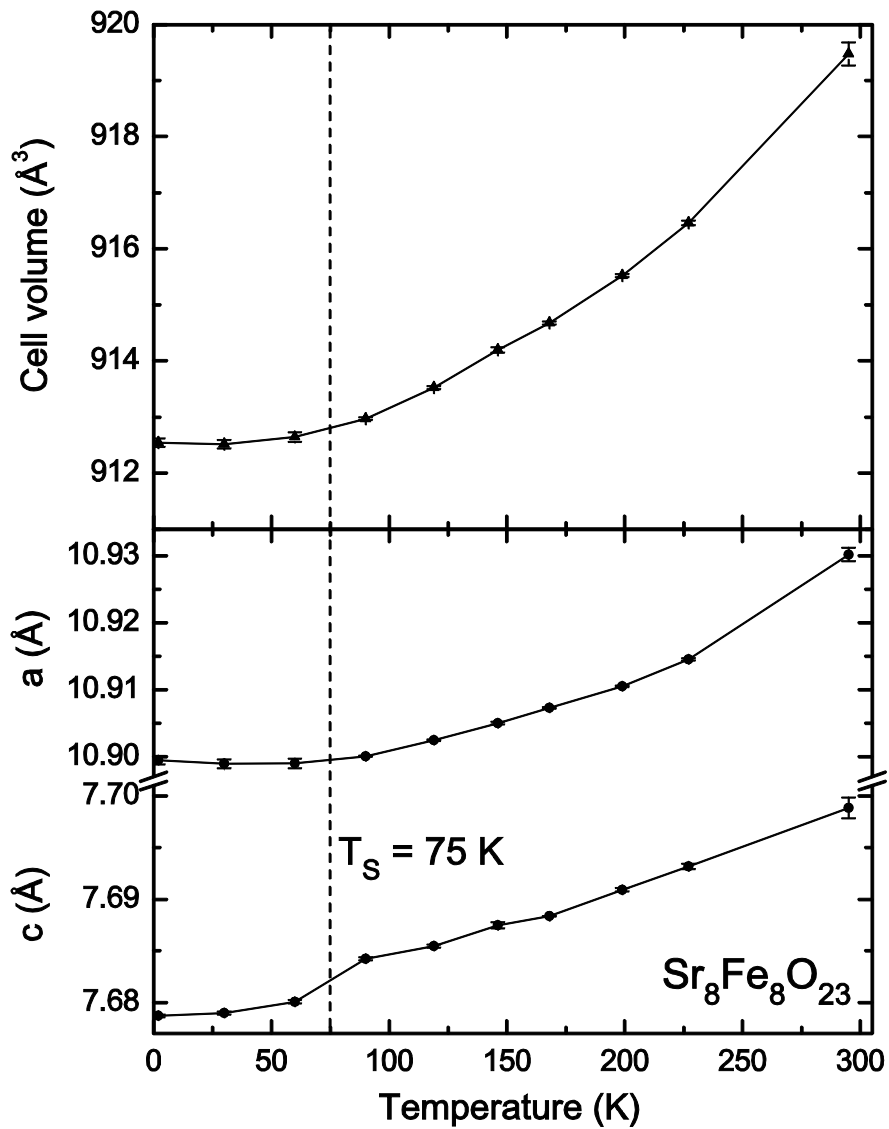

FIG. 9. Temperature dependence of the lattice parameters of $Sr_8Fe_8O_{23}$. With decreasing temperature the tetragonal structure of $Sr_8Fe_8O_{23}$ changes to a monoclinic structure at the transition temperature $T_S = 75(2)$ K. A significant decrease could be observed for the $c$-parameter, while the $a$-parameter shows only a slight increase below 30 K. The $b$-parameter (monoclinic axis) does not show any significant change in the monoclinic phase between 2 and 60 K.

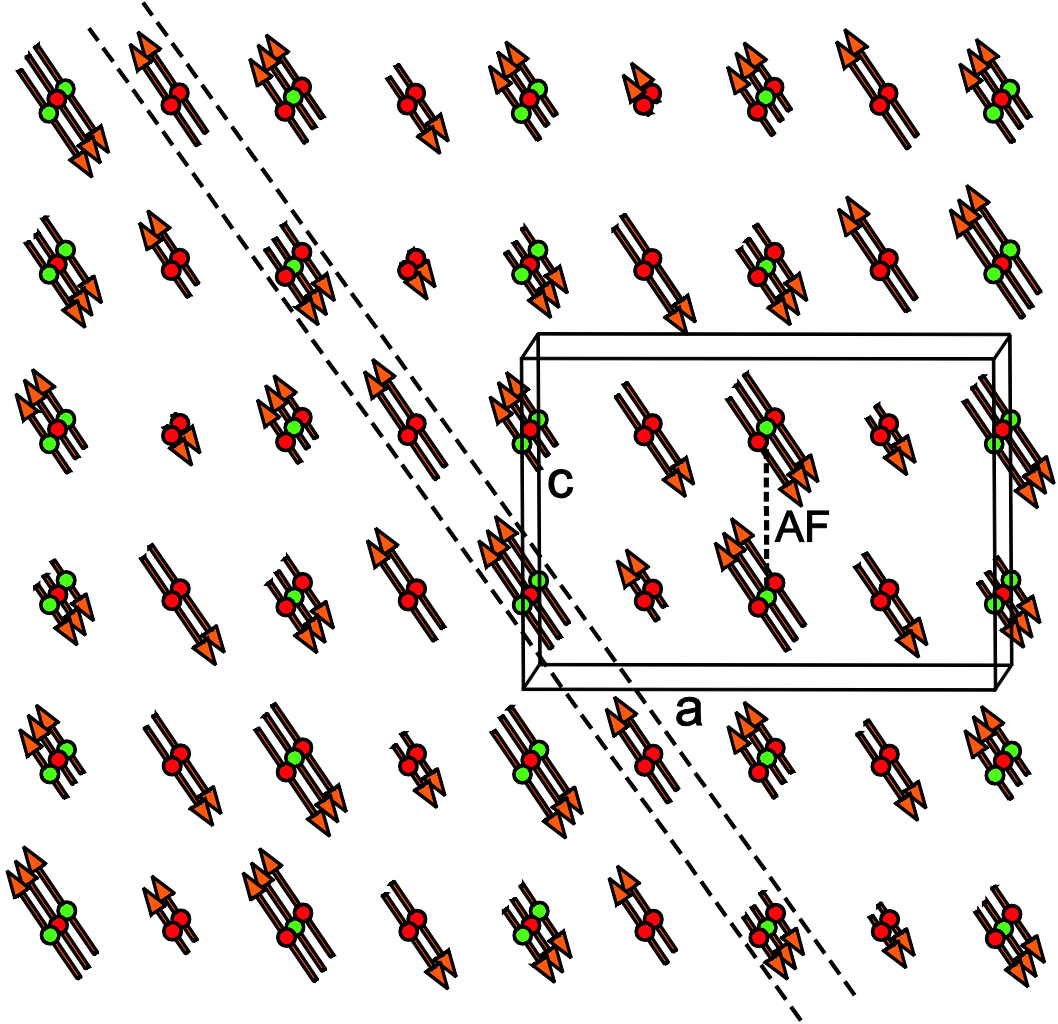

FIG. 10 (Color online). Incommensurate magnetic structure of phase III with the propagation vector $\mathbf{k} = (0.687, 0, 0.326)$, which can be attributed to the magnetic ordering in monoclinic $Sr_8Fe_8O_{23}$. As found earlier for $SrFeO_3$ the magnetic moments of the iron atoms are tilted with respect to the $c$-axis with by angle of $-35.3°$. Thus the moments are aligned within a plane (marked with dashed lines) which is lying perpendicular to the direction [111] using the cubic setting. Due to the oxygen deficiency antiferromagnetic coupling occurs between those iron atoms along the $c$-axis, where sequences of alternating units of octahedral $Fe^{4+}O_6$ and square-pyramidal $Fe^{3+}O_5$ were found ($Fe^{4+}$- and $Fe^{3+}$-ions are marked in red and green, respectively).

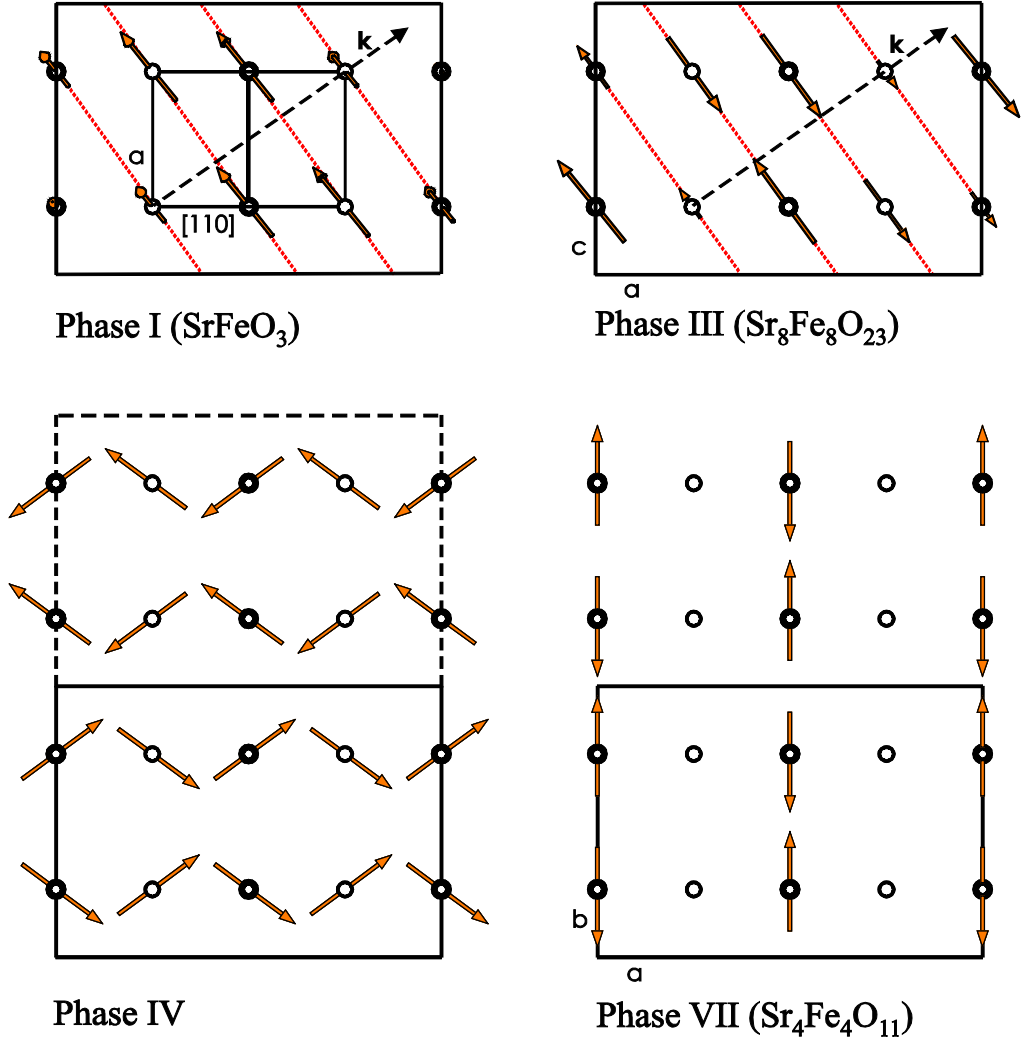

FIG. 11 (Color online). Magnetic ordering of the iron moments in the system $SrFeO_{3-\delta}$. The magnetic structures are shown as projections along the $b$- and $c$-axes of $Sr_8Fe_8O_{23}$ and $Sr_4Fe_4O_{11}$, respectively. For $SrFeO_3$ the cubic unit cell is also shown. The iron moments in $SrFeO_3$ and $Sr_8Fe_8O_{23}$ show a helical and a sine-wave modulated magnetic ordering with a propagation vector $\boldsymbol{k}$ parallel to the direction [111] of the cubic unit cell, respectively. The canting angle of the moments with respect to the $c$-axis is $\theta = -35.3°$. A further decrease of oxygen content leads to commensurate magnetic ordering, where the moments are still canted with respect to the $c$-axis. For this phase the magnetic structure can be described with the propagation vector $\boldsymbol{k} = (0, 0, ½)$ using the tetragonal setting of $Sr_8Fe_8O_{23}$. Thus the magnetic

structure requires a doubling of the tetragonal *c*-axis. A further reduction of oxygen leads to a simple collinear structure of the $Fe^{3+}$-sublattice of $Sr_4Fe_4O_{11}$, whereas the moments of the $Fe^{4+}$-ions in this phase are not three-dimensionally ordered.[8]